\documentclass[aps,prb,twocolumn,groupedaddress,showpacs]{revtex4}

\usepackage{amsmath}
\usepackage{amssymb}
 \usepackage{graphicx}
 \newcommand{\eq}[1]{Eq.~(\ref{#1})}

 \newcommand{\vect}[1]{\mathbf{#1}}
 \newcommand{\uvect}[1]{\hat{\mathbf{#1}}}
 \newcommand{\matr}[1]{\mathbf{#1}}

\newcommand{\Reals}{\mathbb{R}}     

\newcommand{\f}{\ensuremath{\vect{f}}}
\newcommand{\x}{\ensuremath{\vect{x}}}
\newcommand{\Y}{\ensuremath{\matr{Y}}}

\newcommand{\J}{\ensuremath{\matr{J}}}
\newcommand{\eh}{\ensuremath{\uvect{e}}}
\renewcommand{\v}{\ensuremath{\vect{v}}}
\newcommand{\Pmat}{\ensuremath{\matr{P}}}
\newcommand{\oh}{\ensuremath{\uvect{o}}}
\newcommand{\Qmat}{\ensuremath{\matr{Q}}}
\newcommand{\Rmat}{\ensuremath{\matr{R}}}
\newcommand{\breite}{1}

\begin{document}


\title{Experiments with a Malkus-Lorenz water wheel: Chaos and Synchronization}



\author{Lucas Illing}
\email[]{illing@reed.edu}
\author{Rachel F. Fordyce}
\author{Alison M. Saunders}
\author{Robert Ormond}
\affiliation{Physics Department, Reed College, Portland, Oregon, 97202, USA}

\date{\today}
\begin{abstract}
We describe a simple experimental implementation of the Malkus-Lorenz water wheel. We demonstrate that both chaotic and periodic behavior is found as wheel parameters are changed in agreement with predictions from the Lorenz model. We furthermore show that when the measured angular velocity of our water wheel is used as an input signal to a computer model implementing the Lorenz equations, high quality chaos synchronization of the model and the water wheel is achieved. This indicates that the Lorenz equations provide a good description of the water wheel dynamics. 
\end{abstract}



\maketitle

\section{Introduction}

Since the discovery by Lorenz\cite{Lorenz1963} that a simple three-variable set of ordinary differential equations can give rise to exceedingly complex behavior, the study of chaos has continued unabated. Over the years, chaos has been found in a variety of naturally occurring systems and many fascinating table top experiments have been conducted in order to quantitatively study aspects of chaos. Experiments range from dripping faucets~\cite{Dreyer1991} and pendula\cite{Corron2000} to chemical reactions\cite{Schmitz1977} and lasers.\cite{Illing2007} 

For purposes of practical applications of chaos, optical and electronic systems operating on time-scales of nanoseconds or less are a current research focus.\cite{Illing2007,Argyris2005,Cavalcante2010,Callan2010} Yet for purposes of gaining intuition, chaotic mechanical systems operating on timescales of seconds are unsurpassed because of their palpable mechanisms and the direct experience they provide. A prime example of such an experiment is the Malkus-Lorenz water wheel, slowly rotating one way and then another in an at once calming yet interestingly unpredictable fashion.

This water wheel was first envisioned and constructed by Malkus and coworkers as a mechanical analogue of the Lorenz equations.\cite{Lorenz1993,Malkus1972} It has fascinated many students (and teachers) and has become particularly well known since its discussion in Steven Strogatz's introductory text on nonlinear dynamics.\cite{Strogatz1994} The Malkus water wheel described in this paper also grew from such inspiration, starting as a senior thesis project of one of the authors (Rachel Fordyce).  

The most basic Malkus water wheels consist of a few leaking cups attached to the rim of a freely turning wheel whose axis may be either horizontal or tilted. A single stream of water at the top of the wheel will add water whenever a cup  is close to the top. Although the motion of such simple water wheels can be analyzed,\cite{Tylee1995} their dynamics differs from that of the Lorenz model, and we refer to them as non-ideal. The goal in our experiment was to construct an ideal water wheel, one whose dynamics is described by the Lorenz equations.

In this paper we aim to evaluate how close we have come to this goal. We present in Sec.~\ref{sec:setup} details about our experiment, with a focus on confirming three crucial assumptions that are made in deriving the Lorenz model for the wheel's dynamics. The derivation of the model and a brief overview of some of its properties are presented in Sec.~\ref{sec:model}. Experiments resulting in both chaos and periodic oscillations are shown in Sec.~\ref{sec:experiment}. In an effort to make a more precise statement regarding the match of model and experiment, we turn in Sec.~\ref{sec:sync} to an analytic proof and experimental demonstration of chaos synchronization. We conclude with a discussion in Sec.~\ref{sec:discussion}.

\section{Experimental Setup \label{sec:setup}}

\begin{figure}
\centering
\includegraphics[width=\breite \columnwidth]{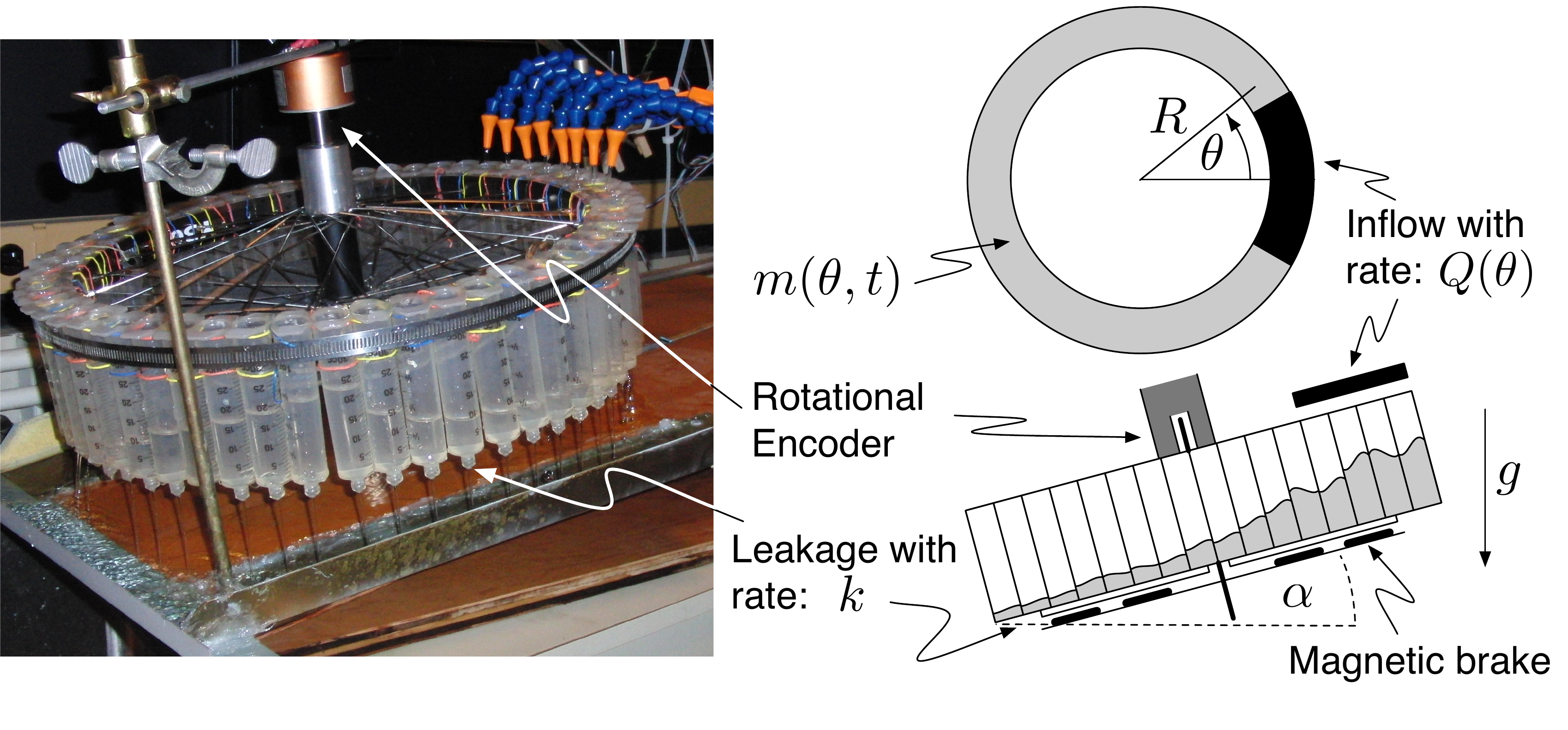}
\caption{(Color online) Experimental implementation of the Malkus-Lorenz water wheel, constructed using a bicycle wheel, syringes, and rare earth magnets forming a magnetic brake. The schematics on the right show a side view (bottom) and a top view (top) of the wheel. (See text for details.) }
\label{fig:setup}
\end{figure}

In designing our water wheel we strove to balance the desire for a simple and inexpensive experiment and our goal to come as close as possible to an ideal water wheel. Three properties of an ideal water wheel require consideration. First, the amount of water entering the cups per unit time has to be constant, which means that gaps between the cups and cup overflow have to be avoided. Second, damping is assumed to be entirely due to frictional torque proportional to velocity (viscous damping), which means that dry friction (kinetic and static friction) has to be negligible in the experiment. 
And third,  the cup leakages should be proportional to the water mass in a cup, which suggests that the outflowing water should exit through a pipe with a laminar flow described by the Hagen-Poiseuille equation.
With these requirements in mind, a water wheel was constructed as described below. 

As a wheel, a smallish 16.5 inch BMX bicycle wheel was used. It was chosen for its smooth bearings that guaranteed  very low axle friction. The wheel was  mounted  to a pair of hinged boards, enabling us to vary the inclination angle $\alpha$ of the wheel with respect to the horizontal (see Fig.~\ref{fig:setup}).

Fifty-six  syringes of 50 cm$^3$ volume each served as leaking ``cups" and were attached to the perimeter of the wheel using a metal synching band. For each syringe, the sides of the finger pull was shaved off in order to allow a snug fit with no gap in between the syringes.  Subsequently to attaching the syringes, the wheel was balanced with small weights, such that the angle at which an empty tilted wheel would come to a stop was distributed uniformly.  We found that rebalancing was necessary in order to remove  the wheel's tendency to preferentially stop with its welding seam at the bottom. 

The motion of the wheel was detected by a rotary encoder attached to the wheel's hub. Its output pulses were counted continuously by an interface circuit containing a HCTL-2016 quadrature decoder and counter chip that includes a 16-bit memory for storage. The counts on the memory chip were then acquired by a computer, with a 10 Hz sampling rate for the data shown in this paper.

To achieve viscous damping,  a (nonmagnetic) Aluminum disc was attached below the wheel such that it co-rotated with the wheel. A second non-rotating disc with several rare earth magnets glued to its surface was mounted opposite the first  disc. The result of this arrangement is a magnetic brake. When the wheel is in motion, the stationary magnets induce eddy currents in the spinning disc which in turn produce magnetic fields that oppose  changes in magnetic flux. A frictional type torque develops that is directly proportional to the angular velocity.\cite{Wiederick1987,Heald1988,Marcuso1991a}
The gap between the discs, and therefore the braking coefficient, is tuneable in our setup.

To check whether damping in our wheel can be modeled as torque linear in the velocity, we measured the angle $\theta$ as a function of time for a slowing wheel without water and fitted the results.
To obtain a model for the fit, we assume that the only relevant torques are from viscous friction due to the magnets and kinetic friction. The magnetic brake contributes a torque $\boldsymbol{\tau}_{\text{vf}}=-\kappa \, \boldsymbol{\omega}$, where $\kappa$ is the viscous friction parameter and $\boldsymbol{\omega}$ is the angular velocity.\footnote{Since one can assume the air drag to be linear in the angular velocity, it is included in this analysis and contributes to $\kappa$.}
 The magnitude of the kinetic friction force is ${F}_{\text{kf}} = \mu \, {F}_{\text{N}}$, where $\mu$ is the friction coefficient and ${F}_N$ is the normal force. Since ${F}_N$ can be taken as constant and the setup is rotationally symmetric, the torque due to kinetic friction is a constant ($|\boldsymbol{\tau}_{\text{kf}}|=\tau_{\text{kf}}=\text{const.}$) opposing the motion of the wheel. 
The evolution of the magnitude of the angular velocity is therefore described by the first-order ordinary differential equation
\begin{align}\label{omegabrake1}
I_{\text{wh}} \, \frac{d {\omega}}{dt} = -\kappa \, \omega - \text{sgn}(\omega) \; \tau_{\text{kf}}, 
\end{align}
where $I_\text{wh}$ is the moment of inertia of the wheel with the attached empty syringes.
Equation~(\ref{omegabrake1}) can be integrated, without loss of generality, under the assumption that $\omega(t=0)=\omega_0 > 0$, yielding
\begin{align}\label{omegabrake}
\omega(t) &= - \frac{\tau_{\text{kf}}}{\kappa} + \left(\omega_0 + \frac{\tau_{kf}}{\kappa} \right) e^{- ({\kappa}/{I_{\text{wh}}})\, t}, &&  t \le t_\text{stop}.
\end{align}
 The magnitude of the angular velocity decreases in time and, once it reaches zero at time $t=t_\text{stop}$, the wheel stops because the torques due to viscous and kinetic friction are zero and static friction implies that some finite minimal torque has to be applied to set the wheel in motion again. 
Integration of \eq{omegabrake}  yields the experimentally measurable angle $\theta(t)$ as a function of time ($t \le t_{\text{stop}}$), 
\begin{align}
\theta 
& = - \frac{ \tau_{\text{kf}}}{\kappa} \, t +  \frac{I_{\text{wh}} }{\kappa} \left( \omega_0 + \frac{\tau_{\text{kf}}}{\kappa}\right)  \left( 1 - e^{- ({\kappa}/{I_{\text{wh}}})\, t} \right),
\end{align}
where we have taken $\theta(0) = 0$. A convenient parameterization is achieved by introducing the viscous damping rate $\gamma = {\kappa}/{I_{\text{wh}}}$ and parameter $\Omega =  \tau_\text{kf}/{\kappa}$, yielding
\begin{align}\label{thetabrake}
\theta  & = - \Omega \, t +  \frac{\omega_0 + \Omega}{\gamma} \left( 1 - e^{- \gamma \, t} \right).
\end{align}

\begin{figure}
\centering
\includegraphics[width=\breite \columnwidth]{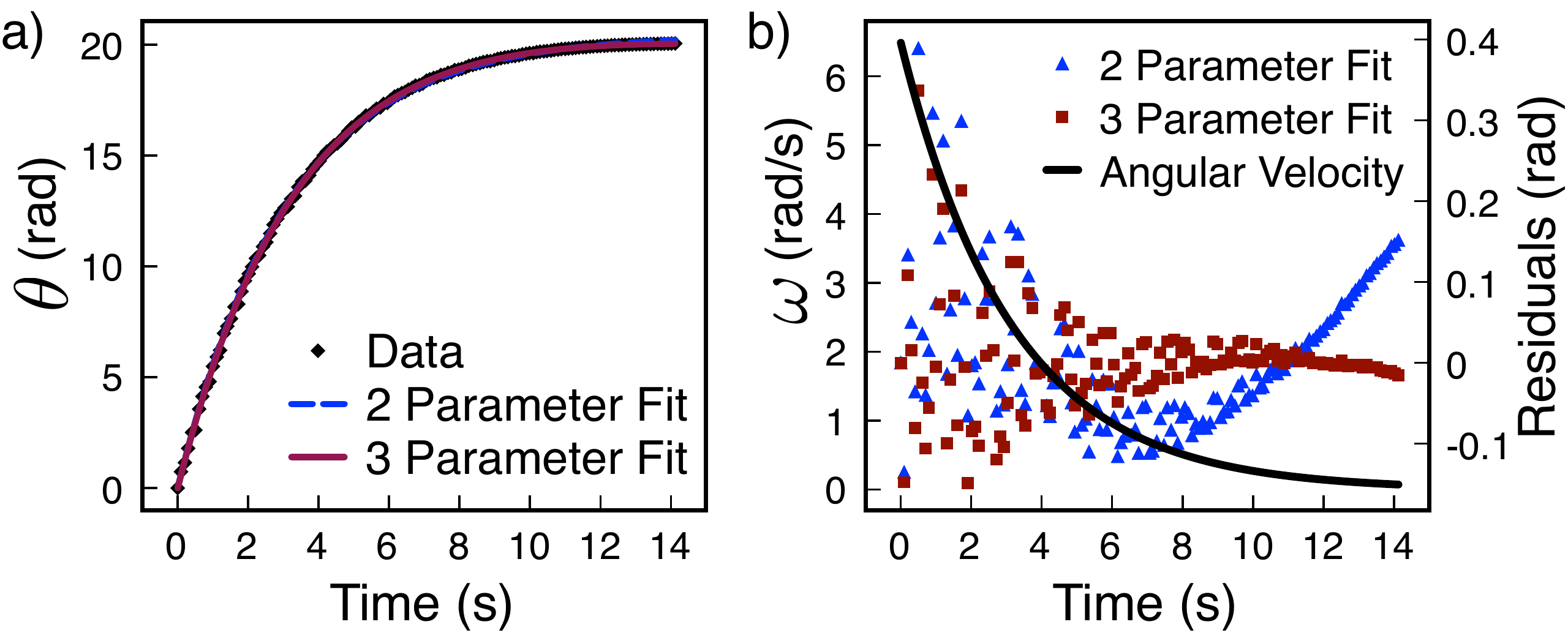}
\caption{(Color online) Measurements of angle versus time for an empty wheel are compared to fits that take into account damping due to viscous friction (2 parameter fit) and a combination of viscous and kinetic friction (3 parameter fit). {(a)} Angle-data and fits. {(b)} Residuals of fits and angular velocity $\omega$ as determined from the 2 parameter fit.  }
\label{fig:magbrake}
\end{figure}

In Fig.~\ref{fig:magbrake} we show two fits of the data to \eq{thetabrake}: a three-parameter fit corresponding to \eq{thetabrake} and a two-parameter fit that corresponds to \eq{thetabrake}  with damping entirely due to viscous friction, that is with $\tau_{\text{kf}}= \Omega =0$. It is seen in Fig.~\ref{fig:magbrake}a that both fits provide an excellent description of the data and that they are essentially indistinguishable to the eye.  The effect of kinetic friction becomes only noticeable when the wheel is turning very slowly.
This is seen from the residuals shown in  Fig.~\ref{fig:magbrake}b, which scatter around a value of zero for the first four seconds when the angular velocity is large, indicating that both models provide a good fit, but show a non-random deviation from zero at later times when the angular velocity is small. In particular, the residuals of the last two seconds before the wheel comes to a stop show that the model with just viscous friction (two-parameter fit) sightly overestimates the angle. 
Despite this, these results clearly indicate that damping is dominated by the magnetic brake and that our water wheel comes close to the ideal one where damping is entirely due to viscous friction. To keep the dynamic model simple, we therefore neglect kinetic friction, from now on, and set $\tau_{\text{kf}}=0$.

The moment of inertia of the wheel can be determined by performing a second measurement and fit for a wheel with a known set of additional weights of total mass $m$ attached symmetrically to the wheel's rim.\cite{Matson2007}   Using the resulting fitted parameter $\tilde{\gamma} = {\kappa}/{(I_{\text{wh}}+m \,R^2)}$, the previously determined $\gamma$, and 
\begin{align}
I_\text{wh} = m R^2 \frac{\tilde{\gamma}}{\gamma-\tilde{\gamma}},
\end{align}
 we found that our wheel has a moment of inertia of $I_\text{wh} \approx 0.11$~kg m$^2$.

For an ideal water wheel it is important that the inflow is symmetric with respect to the centerline that divides the wheel into left and right. In order to achieve this symmetry and, at the same time, minimize the probability of cup overflow while still being able to achieve sufficient inflow rates, we used a system of eight individually adjustable spigots.\footnote{We use Loc-Line hoses and nozzles.} The spigots were placed symmetrically with respect to the top, spanning angles $\theta \in [-\theta_0,+\theta_0]$ with $\theta_0 \approx 26^{\circ}$. A uniform distribution of mass flux, described by 
\begin{align}\label{Qoftheta}
{Q}(\theta) = \begin{cases}
0 & \theta \in [-\pi,\theta_0) \\
\frac{{Q}_{\text{tot}} }{2 \, \theta_0} & \theta \in [-\theta_0,\theta_0]\\
0 &  \theta \in (\theta_0,\pi].
\end{cases},
\end{align}
was ensured by verifying experimentally that for a stationary wheel all eight top cups fill at identical rates. 
The total mass flux $Q_\text{tot}$ was measured using a digital flow meter\footnote{DigiFlow 8000T.} and was tunable with a range of 0.03 - 0.09~kg/s.

\begin{figure}
\centering
\includegraphics[width= \breite \columnwidth]{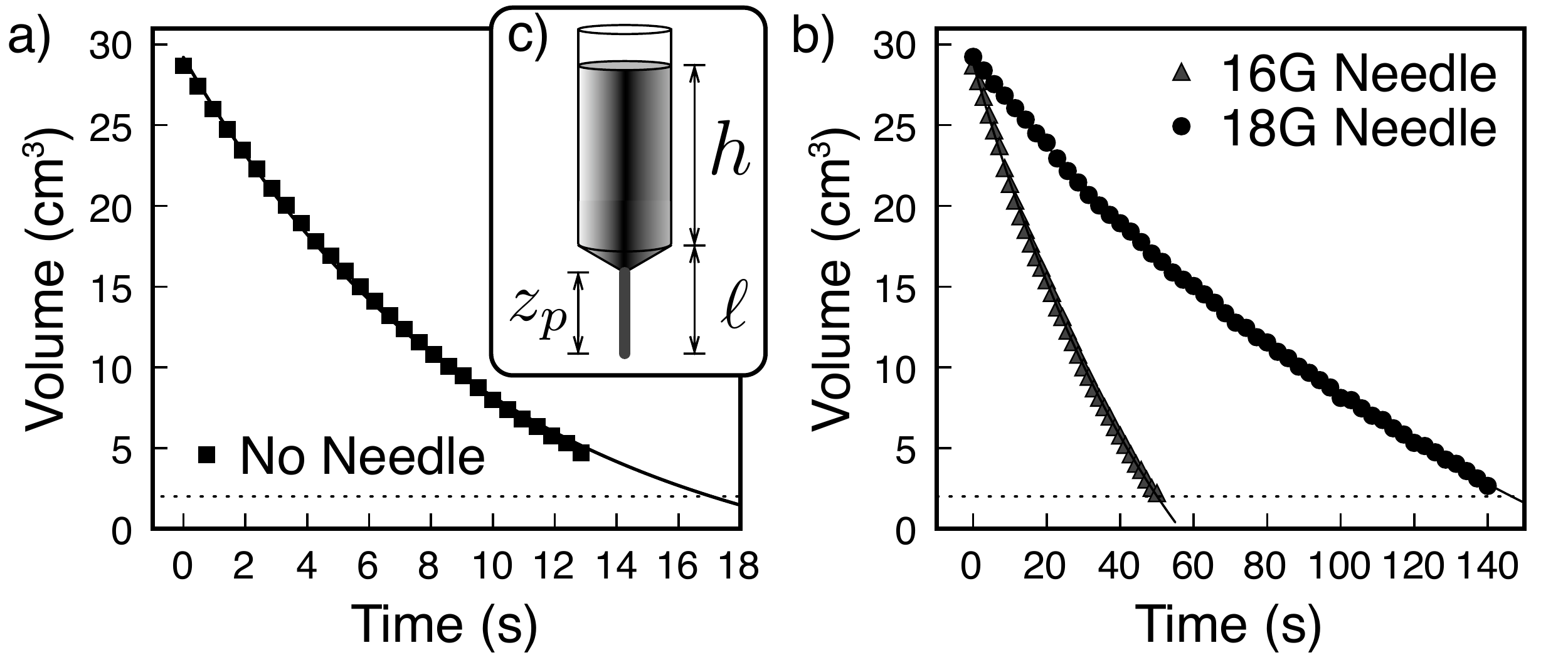}
\caption{Measurements of the water volume in the syringe as a function of time and the two parameter fit of \eq{laminar}. The dotted line indicates the value of $V_b$ (see text). {(a)} Syringe without needle ($V_\text{off}=4$ cm$^3$).  Fit-result: $V_0=29$ cm$^3$, $k=0.10$ s$^{-1}$ {(b)} Syringe with needles ($V_\text{off}=18$ cm$^3$) of 16 gauge  (Fit-result: $V_0=29$ cm$^3$, $k=0.017$ s$^{-1}$) and 18 gauge (Fit-result: $V_0=29$ cm$^3$, $k=0.006$ s$^{-1}$). {(c)} Geometry of the syringe, consisting of a cylindrical main body (radius $r_s$, height $h$), a tapered section, and either a nozzle or a needle as exit ``pipe" (radius $r_p$, height $z_p$). The total height of the tapered section and exit ``pipe" is $\ell$. 
}
\label{fig:leakage}
\end{figure}

Finally, the ideal wheel should have a leakage rate that is proportional to the water mass.
Due to the fact that we use syringes as ``cups", water leaks out either through the small cylindrical nozzles at the end of the syringes, which suggests a theoretical description in terms of a laminar pipe flow of a viscous fluid through the nozzle, or alternatively, needles can be attached resulting in an even longer ``pipe".
Assuming that the laminar pipe flow of the exiting water is the dominant effect, one expects the total volume $V$ of water in the syringe (radius $r_\text{s}=1.08$~cm)  to change according to the Hagen-Poiseuille equation\cite{Pnueli1992}
\begin{align} \label{Poiseuille}
 \frac{dV}{dt} = - \frac{\pi r_\text{p}^{4}}{8 \nu \rho_\text{w} } \left( \frac{ \Delta P}{z_p}\right). 
 \end{align}
 Here, $r_\text{p}$ is the exit ``pipe" radius,  $\rho_\text{w}$  is the density, and $\nu$ the kinematic viscosity of water. The  change  $\Delta P$ of the modified pressure $P$ over the exit pipe length $z_p$ is assumed to be entirely due to gravity, implying $\Delta P=g \,\rho_\text{w} \, (h+\ell)$  (see Fig.~\ref{fig:leakage}c). The total volume $V=\pi r_\text{s}^2 h + V_\text{b}$ consists of the water in the cylindrical part of the syringe and an additional approximately 2 cm$^3$ volume $V_\text{b}$ at the bottom. Taken together, this suggests that the volume in a syringe behaves as  
 \begin{align} \label{Poiseuille2}
 \frac{dV}{dt} = - k \left( V  + V_\text{off} \right) 
 \end{align}
 where $V_\text{off} = \ell \pi r_\text{s}^2 - V_\text{b}$ is  a constant offset term and the rate $k$ is a parameter that we fit (nominally  $k =  r_\text{p}^{4} \, g/8 \, \nu \, r_\text{s}^2 \, z_p$). The fit model is obtained by integrating \eq{Poiseuille2}, yielding
 \begin{align}\label{laminar}
V(t) = V_0 \, e^{- k \, t}  + V_\text{off} \, (e^{- k \, t}-1).
\end{align}
This model is valid for water levels within the cylindrical part of the syringe, \emph{i.e.} for $V(t)>V_\text{b}$. The drainage of the small bottom volume cannot be measured with our setup but we is expect it to deviate somewhat from model~(\ref{laminar}) due to the tapering of the syringe and capillary effects. However, since the bottom volume is very small, we neglect any potential small corrections and assume that \eq{laminar} describes the leakage of the entire syringe.

To check the validity of \eq{laminar}, we determined the total water volume as a function of time by using motion tracking software to analyze digital films of the declining water level in a vertical syringe with and without attached needles. The data and fit results are shown in Fig.~\ref{fig:leakage}.  It is seen that the fit is quite accurate already for the syringe without needle, as shown in Fig.~\ref{fig:leakage}a, and becomes even better when a needle is attached, as seen in Fig.~\ref{fig:leakage}b. Note that by attaching needles and by changing the gauge of the needles the leakage rate $k$ can be varied by orders of magnitude.  No needles were attached to the wheel for all the  data shown in this paper.

Writing \eq{Poiseuille2} in terms of the total mass of water in all 56 syringes, $M_\text{tot} = 56 \rho_\text{w} \, V$, and introducing an effective mass flux $Q_\text{eff}$ via 
\begin{align}\label{Qeff}
Q_\text{eff}(\theta) ={Q}(\theta) - 56/(2\pi) \, k \, \rho_\text{w} \, V_\text{off},
\end{align}
where $Q(\theta)$ is given by \eq{Qoftheta}, one finds that  the total mass in the wheel evolves according to 
\begin{align}\label{totmass}
\frac{d M_{\text{tot}}}{dt} = -k \, M_{\text{tot}} + Q_{\text{eff}}^\text{tot},
\end{align}
with $Q_\text{eff}^\text{tot}  = Q_\text{tot} - 56 \, k \, \rho_\text{w} \, V_\text{off}$. Equation (\ref{totmass})
shows that the total mass of water is conserved in the asymptotic limit because $M_{\text{tot}} = Q_{\text{eff}}^\text{tot}/k$ for $t \to \infty$.

\section{The Model \label{sec:model}}

To make this paper self-contained, we recall here the well known derivation of the equations describing the dynamics of the water wheel and briefly discuss the parameter dependence of the model solutions.

\subsection{Derivation}

In our derivation we follow \citet{Strogatz1994} and model the water as being distributed as a continuous ring around the rim of the wheel. \citet{Matson2007} showed that the same result can be obtained when considering an ideal water wheel with a discrete set of cups.

 In the continuum approximation the mass distribution $m(\theta,t)$ around the wheel's rim is defined such that the mass between the angles $\theta_1$ and $\theta_2$ (fixed in the lab frame) is 
\begin{align}
M(t) = \int_{\theta_1}^{\theta_2} m(\theta,t) \, d\theta.
\end{align}
The change in time of $m(\theta,t)$, given by 
\begin{align}\label{meq}
\frac{\partial m(\theta,t)}{\partial t}  = Q_\text{eff}(\theta) - k \, m(\theta,t) - \omega(t) \, \frac{\partial m(\theta,t)}{\partial \theta},
\end{align} 
 has three contributions: the constant inflow $Q(\theta)$, described by \eq{Qoftheta}; the leakage, which is proportional to the water mass and which also contributes a constant offset to the inflow, resulting in $Q_\text{eff}$ as described by \eq{Qeff}; and a third term that takes into account the wheel's rotation.
The third term expresses the fact that the mass density at angle $\theta$ and time $t+\Delta t$, \emph{i.e.} $m(\theta,t+\Delta t)$, is identical to the mass density at time $t$ at angle $\theta - \Delta \theta$, \emph{i.e.} $m(\theta- \Delta \theta,t)$, for a wheel without water inflow and leakage, rotating with angular velocity $\omega$ such that $\Delta \theta = \omega(t) \, \Delta t$.

The change in time of the angular velocity is due to the applied total torque, which has three components. The magnetic brake contributes $\tau_{\text{vf}} = - \kappa \, \omega(t)$, as discussed in Sec.~\ref{sec:setup}. The infalling water contributes a torque $\tau_{\text{spin up}} = - Q_{\text{tot}}\, R^2 \, \omega(t)$ because it enters the wheel a distance $R$ from the center with zero angular velocity and is spun up to an angular velocity of $\omega$ before leaking out. Gravity provides the third torque  because each infinitesimal mass element $m(\theta,t) \, d\theta$ on a wheel that is tilted at an angle $\alpha$ contributes
\begin{align}
d\tau_{\text{grav}} =  R \, g \, \sin( \alpha) \, \sin( \theta) \, m(\theta,t) \, d\theta.
\end{align}
Taken together, one obtains 
\begin{align}\label{omeq}
I_{\text{tot}} \, \frac{d \omega(t)}{dt} =& - \left( \kappa + Q_{\text{tot}}\, R^2 \right) \, \omega(t) \nonumber \\
&+  R \, g \, \sin( \alpha) \, \int_{-\pi}^{\pi} m(\theta,t)  \sin( \theta) \, d\theta,
\end{align}
where we have taken $I_{\text{tot}}$ to be a constant, which is valid after an initial transient, \emph{i.e.} after  the total water mass has come exponentially close to its final constant value [see \eq{totmass}].

Since $m(\theta,t)$ is periodic in $\theta$, one can expand this function into Fourier-modes,
\begin{align}
m(\theta,t) &= \sum_{n=1}^{\infty} a_n(t) \,  \sin( n \, \theta) +  \frac{b_0(t)}{2} + \sum_{n=1}^{\infty} b_n(t) \, \cos( n \, \theta).
\end{align}
Under the assumption that $Q(\theta)$ is truly symmetric with respect to the wheel's center line, the effective inflow  $Q_\text{eff}$ can be written as
\begin{align}
Q_\text{eff}(\theta) &=  \frac{q_0}{2} + \sum_{n=0}^{\infty}  q_n \, \cos( n \, \theta) - \frac{56}{2 \pi} k \rho_\text{w} V_\text{off}, 
\end{align}
with  coefficients
\begin{align}\label{qfac}
q_n &=  \frac{Q_{\text{tot}} }{\pi} \, \text{sinc}(n \theta_0) &(n&=0,1,2,\ldots),
\end{align}
for the case that $Q(\theta)$ is given by \eq{Qoftheta}.
Substituting these series into the partial differential equation~(\ref{meq}) and integro-differential equation~(\ref{omeq}) and equating the coefficients of each harmonic separately, one obtains an infinite set of ordinary differential equations that describes the evolution of the angular velocity $\omega(t)$ and the Fourier-mode amplitudes $a_n(t)$ and $b_n(t)$. Amazingly, the evolution of the angular velocity and amplitudes $a_1(t)$ and $b_1(t)$ is entirely decoupled from the evolution of all other amplitudes. Therefore, the problem is reduced to a three-dimensional system of ordinary-differential equations
\begin{align}\label{origlor}
\begin{split}
\frac{d \omega}{dt} &= - \frac{\kappa + Q_{\text{tot}}\, R^2}{I_{\text{tot}}} \,  \omega(t) + \frac{\pi \, R \, g \, \sin \alpha}{I_{\text{tot}}} \, a_1(t) \\
\frac{d a_1}{dt} &= - k \, a_1(t) +  \omega(t) \, b_1(t) \\
\frac{d b_1}{dt} &= q_1 - k \, b_1(t)  -  \omega(t) \, a_1(t)  \\
\end{split}
\end{align}
It turns out that \eq{origlor} can be mapped onto the Lorenz equations by introducing the dimensionless parameters
\begin{align}
\sigma  &= \frac{1}{k} \frac{\kappa + Q_{\text{tot}}\, R^2}{I_{\text{tot}}}, &
\rho &= \frac{q_1}{k^2} \frac{\pi \, R \, g \, \sin \alpha}{\kappa + Q_{\text{tot}}\, R^2}, 
\end{align}
the dimensionless time $s=k t$, and the coordinates
\begin{align}\label{coordtrafo}
x &= \frac{\omega}{k}  &  y &= \frac{\rho \, k}{q_1} \, a_1 & z&= \rho - \frac{\rho \, k}{q_1} \, b_1.
\end{align}
In the new coordinates, \eq{origlor} becomes 
\begin{align}\label{lor}
\begin{split}
\dot{x} &= \sigma \, (y-x) \\
\dot{y} &= \rho \, x - y - x \, z \\
\dot{z} &= x \, y - z
\end{split},
\end{align}
where the overdot denotes the derivative with respect to the dimensionless time $s$. Equation~(\ref{lor}) is identical to the Lorenz equations with the third parameter (often denoted as $b$) equal one.

Above mapping of the problem onto Eq.~(\ref{lor}) relies in an essential way on the fact that Eq.~(\ref{meq}) is linear in $m(\theta,t)$, explaining why an ideal water wheel has to have a leakage that is proportional to the water mass. Furthermore, to obtain the correct form of the Lorenz model, Eq.~(\ref{omeq}) needs to be linear with respect to $\omega(t)$, making a purely viscous damping a principal requirement of ideal water wheels.

\subsection{Characterization via Lyapunov Exponents}

Starting with Lorenz's seminal 1963 paper,\cite{Lorenz1963} the Lorenz equations have been studied intensively over the last decades.  What makes them so interesting is that this simple looking deterministic system has extremely rich dynamics.
For some parameters the asymptotic solutions are steady state solutions, for other parameters periodic oscillations are found.\cite{Strogatz1994,Sparrow1982}  In addition, and this is the reason that the Lorenz equations are famous, over wide parameter ranges the solutions exhibit a qualitatively different and new form of behavior, they are chaotic. That is, they oscillate irregularly, never exactly repeating, and depend very sensitively on the choice of initial conditions (the initial values of the three variables). 
Some rigorous results are available on the Lorenz equation's chaotic solutions and bifurcations,\cite{Sparrow1982, Galias1998,Tucker1999,Mischaikow2001} where a bifurcation refers to a sudden qualitative change in behavior that occurs when a small smooth change is made to a parameter value. Yet, since such rigorous results are extremely challenging to obtain, one generally resorts to approximate numerical calculations to characterize solutions.

To numerically classify solutions of \eq{lor} as a function of the parameters, we compute the largest Lyapunov exponents. Lyapunov exponents provide a useful measure because they are independent of the particular initial conditions used in computing them and are invariant  under smooth nonsingular coordinate transformations such as those in \eq{coordtrafo}.
In addition they directly measure a distinguishing characteristic of chaotic solutions, their  ``sensitivity to initial conditions". 
For chaotic systems, small differences in initial conditions yield after an exponentially short time widely diverging solutions, rendering impossible the long-term prediction of outcomes from finite-precision measurements of initial states.
Roughly speaking, two trajectories in phase space with initial separation $\delta \x_0$  diverge as
\begin{align}\label{LErough}
    | \delta \x(t) | \approx e^{\lambda t} | \delta \x_0 |,
\end{align}
and $\lambda$, the largest Lyapunov exponent (LLE), is a positive real number for a chaotic system. 
In contrast, stable steady state solutions have a negative LLE, because perturbations away from the steady state decay exponentially, and stable periodic solutions have a LLE with value zero,  because perturbations along a trajectory neither grow nor shrink in time (and perturbations perpendicular to the trajectory decay exponentially).

\begin{figure}
\centering
\includegraphics[width=\breite \columnwidth]{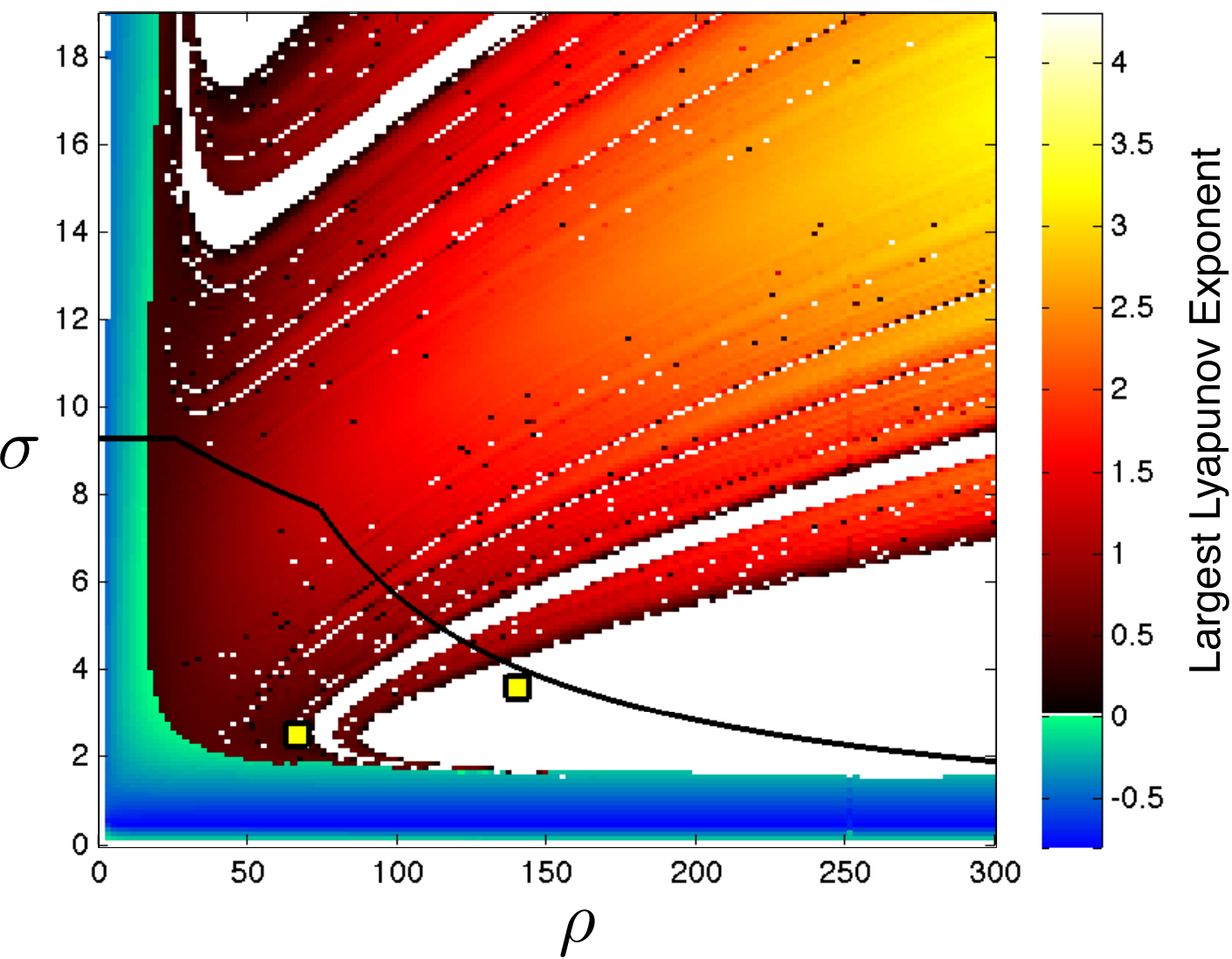}
\caption{ (Color) Shown is the numerically determined largest Lyapunov exponent as a function of $\rho$ and $\sigma$. White corresponds to a zero exponent implying periodic oscillation of $\omega$. Red colors correspond to positive and blue colors to negative exponents, indicating, respectively, chaos and steady state evolution of $\omega$. The approximate upper boundary of the experimentally accessible regions is shown by the solid black line. The yellow squares indicate parameter values corresponding to the chaotic and periodic experimental time series in Fig.~\ref{fig:ts}. }
\label{fig:maxle}
\end{figure}

Figure~\ref{fig:maxle} depicts the results of a numerical calculation of the LLEs, 
providing a map of the water wheel's dynamic behavior as a function of the relevant effective parameters, $\rho$ and $\sigma$. (Details about the numeric calculation are given in the Appendix.)
It is seen that for small values of either parameter only negative LLEs are found, corresponding to steady state behavior of the angular velocity. For larger parameter values, the wheel exhibits either chaotic or periodic behavior. We find that there exist large continuous regions with a zero LLE, shown in white, indicating periodic behavior. In addition, there are large continuous chaotic regions with positive LLEs, shown in red colors (dark gray).  Yet, this numerical calculation also reveals that there are periodic windows embedded inside the chaotic region and that those windows seem to form a fractal set, with an increasing number of ever smaller periodic windows, each organized along a line in parameter space.  This intricate structure means that the qualitative dynamics of the Lorenz system sensitively depends on the chosen parameter values. 
It also should be noted that for some parameter values several different asymptotic solutions (i.e. attractors) may coexist in phase space.\cite{Sparrow1982} Our numeric result only depicts the LLE associated with one of those solutions, the one whose basin of attraction includes the randomly chosen initial condition of the simulation.

\section{Experimental Chaos \label{sec:experiment}}

\begin{figure}
\centering
\includegraphics[width= \breite \columnwidth]{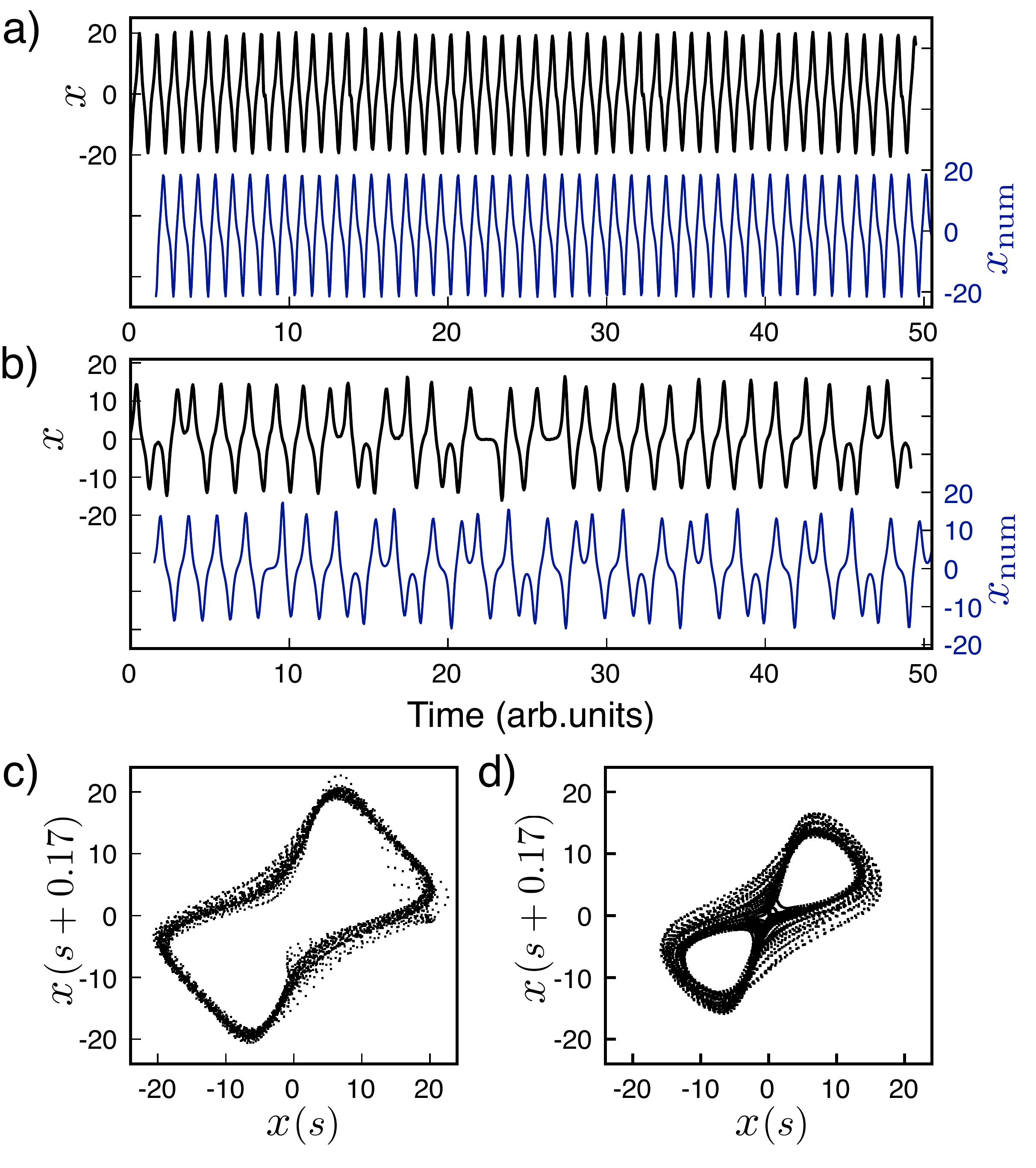}
\caption{(Color online)  {(a,b)} The experimental time series of the rescaled angular velocity, $x= \omega/k$, as a function of the dimensionless time $s = k \,t$ is shown as the upper trace and the corresponding numerical solution $x_\text{num}$ as the lower trace. The parameters are {(a)} $\sigma \approx 3.6$ and $\rho \approx 140$, resulting in periodic oscillations, and {(b)} $\sigma \approx 2.5$ and $\rho \approx 66$, resulting in chaotic oscillations.  A corresponding two dimensional time-delay embedding of the experimental data is shown in {(c)} for the periodic and {(d)} for the chaotic case.  }
\label{fig:ts}
\end{figure}

Figure~\ref{fig:maxle} is a useful guide for experiments because it informs our choice of wheel parameters that result in a desired dynamic behavior. 
This a-priori connection between the wheel operating condition and expected dynamic behavior is possible because we can explicitly write the effective parameters $\rho$ and $\sigma$ as a function of the experimentally tunable and fixed measurable parameters. Leaving aside the option of attaching needles to the wheel, the experimentally adjustable parameters are the inclination angle $\alpha$, the viscous damping rate $\gamma$, and the total mass flux $Q_\text{tot}$. We obtain $\rho$ and $\sigma$ in terms of these quantities by substituting \eq{qfac} for $q_1$, writing $I_\text{tot}$ as  $I_\text{tot} = I_\text{wh} + M_\text{tot} R^2 = I_\text{wh} + Q_\text{eff}^\text{tot} R^2/k$ with $Q_\text{eff}^\text{tot}  = Q_\text{tot} - 56 \, k \, \rho_\text{w} \, V_\text{off}$,  and using $\kappa = \gamma \, I_\text{wh}$. These substitutions yield
\begin{align}
\sigma(Q_\text{tot},\gamma)  &= \frac{\gamma \, I_\text{wh} + Q_{\text{tot}}\, R^2}{k \, I_\text{wh} + Q_{\text{tot}}\, R^2 - 56 \, k \, \rho_\text{w} V_\text{off} \, R^2}, \label{sig} \\
\rho(Q_\text{tot}, \gamma, \alpha) &= \frac{Q_\text{tot} \, R \, g  \, \text{sinc}(\theta_0)  \, \sin(\alpha)}{k^2 \, (\gamma \, I_\text{wh} + Q_\text{tot} \, R^2)}. \label{rho}
\end{align}
One useful fact, made apparent by these equations, is that one can tune the parameter $\rho$ independently of $\sigma$ by varying the wheel's inclination angle $\alpha$.
Furthermore, using these equation, we can determine the portion of parameter space that can be accessed with our implementation of the water wheel. The accessible region's approximate upper boundary is shown in  Fig.~\ref{fig:maxle} by the black solid line. 

Although it is experimentally not possible to resolve the fine fractal structure seen in Fig.~\ref{fig:maxle}  because of experimental limitations such as noise and parameter drift, we do find that our water wheel does qualitatively conform to the theoretical predictions. For small $\rho$ or $\sigma$, it shows steady state behavior of the angular velocity, corresponding to either stationary behavior or rotation with constant speed. For larger parameter values we find chaos or periodic oscillations. Furthermore, for fixed values of $\sigma$ ($\sigma \approx 3$), we find that the wheel is in a chaotic state for small angles $\alpha$ (small $\rho$) and exhibits periodic behavior for larger angles $\alpha$ (large $\rho$).

As an example of the obtainable data, we display in Fig.~\ref{fig:ts} two experimental time series, one periodic (Fig.~\ref{fig:ts}a) and one chaotic (Fig.~\ref{fig:ts}b). For each run several hours of data are taken, recording the wheel's angle $\theta$ as a function of time. The first roughly one hour is discarded as transient and the rest is used for further time series analysis. 
Next, we low-pass filter and take the first derivative of the data in the Fourier-domain. Working in the Fourier domain is an efficient way of suppressing high frequency noise that would otherwise dominate the derivative. It is a legitimate way of proceeding because the data is highly oversampled (on the order of 100 points per oscillation period) and the Fourier spectrum has a dominant (but broadened) peak even for chaotic time series, which enables us to chose a filter that has essentially no effect at dynamically relevant time-scales. For example, the angle data corresponding to the chaotic time series in Fig.~\ref{fig:ts}b is peaked around 0.07~Hz and the filter cutoff was chosen at $\sim0.6$~Hz. Having computed the derivative, i.e. the angular velocity $\omega$, we obtain $x$ by using the coordinate transformation~(\ref{coordtrafo}), $x=\omega/k$.

The upper (black) traces shown in Fig.~\ref{fig:ts} are $\sim$7 min windows of data obtained after such processing. The time trace in Fig.~\ref{fig:ts}a is clearly periodic. The small fluctuations of the amplitude are a result of the unavoidable imperfections of the experiment. These become even more apparent  in the two dimensional time-delay embedding\cite{Abarbanel1995,Kantz2004} of the same data that is shown in Fig.~\ref{fig:ts}c.  For an ideal noise-free experiment exhibiting periodic oscillations, a single closed loop (a limit cycle attractor) should be seen, whereas our actual data does result in a smeared-out loop. Nevertheless, the correspondence between experiment and model is excellent, as can be seen in Fig.~\ref{fig:ts}a by comparing  the upper and the lower trace, which are, respectively, the experimental data and a numerical solution of the Lorenz equations [\eq{lor}].

The time series in Fig.~\ref{fig:ts}b is clearly non-periodic. A time-delay embedding of the same data is depicted in Fig.~\ref{fig:ts}d and shows a complicated (potentially fractal) attractor. Together this indicates that the wheel oscillates chaotically. 
 For comparison, we also display in Fig.~\ref{fig:ts}b the result of a simulation of the Lorenz equations with the corresponding parameter values of $\rho$ and $\sigma$, which lie in the chaotic regime of the model (see square in Fig.~\ref{fig:maxle}). Again, we find a convincing qualitative correspondence between experiment and model in the sense that amplitude, dominant frequency, and general character of the solutions match. Since the solutions are chaotic they are, of course, not expected to be identical.

Thus, qualitatively we find a good correspondence of the observed water wheel behavior and the predictions of the Lorenz equations. In order to obtain a more precise statement about the correspondence of model and experiment, we utilize another fascinating aspect of nonlinear dynamics: chaos synchronization.

\section{Chaos Synchronization\label{sec:sync}}

Though the synchronization of clocks (periodic systems) has been studied
with great care over centuries,\cite{Pikovsky2003} the discovery\footnote{Chaos synchronization was independently discovered three times:  by H. Fujisaka and T. Yamada in Japan [Prog. Theor. Phys. \textbf{69}, 32-47 (1983)], by V. S. Afraimovich, N. N. Verichev, and M. I. Rabinovich in the USSR [Radiophys. Quantum Electron. \textbf{29}, 795-803 (1986)], and by L. M. Pecora and T. L. Carroll in the US [Phys. Rev. Lett. \textbf{64}, 821-824 (1990)].} that two chaotic oscillators could synchronize came as a surprise. Synchronization was unexpected because sensitivity to small perturbations means that the solutions of two chaotic systems tend to diverge.
Yet, it was found that, if two chaotic systems are coupled to each other in a suitable way, it is possible for one chaotic system  to follow exactly the time evolution of an identical second chaotic system, even when the chaotic systems start from very different initial conditions. Indeed, for the Lorenz system one can even prove that this is the case. 

The phenomenon of chaos synchronization suggests the following test of the validity of the Lorenz equations as a description of the water wheel.  Presume that the water wheel dynamics is described by the Lorenz equations. Then a computer model has to synchronize to the water wheel for a suitably chosen coupling (e.g. for sufficient coupling strengths). That is, if we measure the angular velocity of the water wheel as a function of time, thereby determining $x(t)$ in \eq{lor}, and use this data as input to a numerical Lorenz model given by
\begin{align}\label{lorm}
\begin{split}
\dot{x}_\text{m} &= \sigma \, (y_\text{m}-x_\text{m})  - K \, (x_\text{m} - x)\\
\dot{y}_\text{m} &= \rho \, x_\text{m} - y_\text{m} - x_\text{m} \, z_\text{m} \\
\dot{z}_\text{m} &= x_\text{m} \, y_\text{m} - z_\text{m},
\end{split}
\end{align}
where $K$ sets a sufficiently large coupling strength, the model output $x_\text{m}(t)$ has to converge to $x(t)$, \emph{i.e.}  $x_\text{m} \to x$ as $t \to \infty$. If no convergence is observed, then the water wheel dynamics is not accurately modeled by the Lorenz equations.

Before describing how our wheel performs under this test, let us discuss the proof of synchronization.  

\subsection{Proving Synchronization}

First note that exactly synchronized solutions of the coupled system, \eq{lor} and \eq{lorm}, correspond to a zero error-vector, $\mathbf{e}(t)=0$, where the components of  $\mathbf{e}$ are defined as $(e_x,e_y,e_z)=(x_\text{m}-x,y_\text{m}-y,z_\text{m}-z$). It is therefore convenient to consider the evolution of $\mathbf{e}$, which is described by a non-autonomous ordinary differential equation of the form $\dot{\mathbf{e}} = \mathbf{F}(\mathbf{e},t)$, given by
\begin{align}\label{err}
\begin{split}
\dot{e}_x &= \sigma \, (e_y - e_x)  - K \, e_x\\
\dot{e}_y &= \rho \, e_x - e_y - e_x \, e_z - x(t) \, e_z - z(t) \, e_x \\
\dot{e}_z &= e_x \, e_y  - e_z + x(t) \, e_y + y(t) \, e_x.
\end{split}
\end{align}
Solutions $\mathbf{e}(t)$ can be thought of as trajectories in a three-dimensional phase space. 
To prove synchronization, we need to show that for any initial condition the corresponding solution trajectory asymptotically approaches the origin of this phase space, $\mathbf{e} \to 0$ as $t \to \infty$. Since the solutions of the Lorenz equations~(\ref{lor}) enter in \eq{err}, it is useful to establish bounds for the Lorenz solutions. Bounds can be obtained by constructing a trapping region, that is, a closed region in the phase space of \eq{lor} with the property that solutions starting on the  inside of this region never leave it and solutions starting on the outside enter it after some time. For the Lorenz equations with $b=1$ [\eq{lor}] the following inequalities define a trapping region\cite{Swinnerton2001}
\begin{align}\label{bounds}
x^2&<4\, \sigma \, \rho & 
y^2 &\le \rho^2 &
0 &\le z \le 2 \rho,
\end{align}
for  positive real-valued parameters $\sigma$ and $\rho$. 

Next, we utilize a function that decreases along trajectories of system (\ref{err}), a so-called Lyapunov function.  As a Lyapunov function candidate consider the scalar function
\begin{align}
V(\mathbf{e}) = \frac{e_x^2+e_y^2+e_z^2}{2},
\end{align}
which is positive definite for $\mathbf{e} \ne 0$, zero for $\mathbf{e}=0$, and monotonically increasing as a function of $|\mathbf{e}|$.
If we can show that $\dot{V} < 0$ for all $\mathbf{e} \ne 0$, then all trajectories flow ``downhill" toward the origin and $\mathbf{e}=0$ is globally asymptotically stable.\cite{Strogatz1994}  Using \eq{err} we obtain 
\begin{align}
\dot{V}(\mathbf{e}) = & \, e_x \, \dot{e}_x + e_y \, \dot{e}_y  + e_z \, \dot{e}_z \nonumber \\
 = & - \left[ K + \sigma - \frac{(\sigma + \rho - z)^2}{4} - \frac{y^2}{4} \right] \, e_x^2 \nonumber \\
&- \left( e_y - \frac{\sigma + \rho - z}{2} \, e_x \right)^2 - \left( e_z - \frac{y}{2} \, e_x \right)^2 \label{Vdot}
\end{align}
 for the time derivative of $V$.
Utilizing \eq{bounds}, a lower bound for the expression in the square brackets is found  in terms of the time-independent parameters $\rho$ and $\sigma$,
\begin{align}
K + \sigma - \frac{(\sigma + \rho - z)^2}{4} - \frac{y^2}{4} \ge  K + \sigma - \frac{(\sigma + \rho)^2}{4} - \frac{\rho^2}{4} ,
\end{align}
which, in turn, shows that for sufficiently large coupling strengths $K$,
\begin{align}\label{couplstr}
4 K > \sigma^2 + 2 \sigma  \rho + 2  \rho^2 - 4 \sigma,
\end{align}
 the square-bracket term is positive definite. Then it follows from \eq{Vdot} that
\begin{align}
\dot{V}(\mathbf{e}) < 0 \qquad \forall \, \mathbf{e}\ne0.
\end{align}
This concludes the proof. 

It is important to emphasize that above proof and, in particular, condition (\ref{couplstr}) are sufficient conditions that guarantee that two Lorenz systems coupled unidirectionally, as described by \eq{lorm}, will identically synchronize. Generally, the thus obtained value of $K$ is a conservative estimate. Systems often synchronize for much smaller coupling strengths.
It should also be noted that, in the $K \to \infty$ limit, the model output $x_m$ will converge to any given input $x$, independent of the form of the dynamical system that generated $x$. In other words, the convergence of $x_m$ to input $x$ does not prove the correctness of the Lorenz model. However, if $x_m$ does not converge to a (noise-free) input $x$ for the sufficient coupling strength $K$ derived above, $x$ is definitely not generated by a system that evolves according to the Lorenz equations.

\subsection{Synchronizing the Model and the Wheel}

\begin{figure}
\centering
\includegraphics[width= \breite \columnwidth]{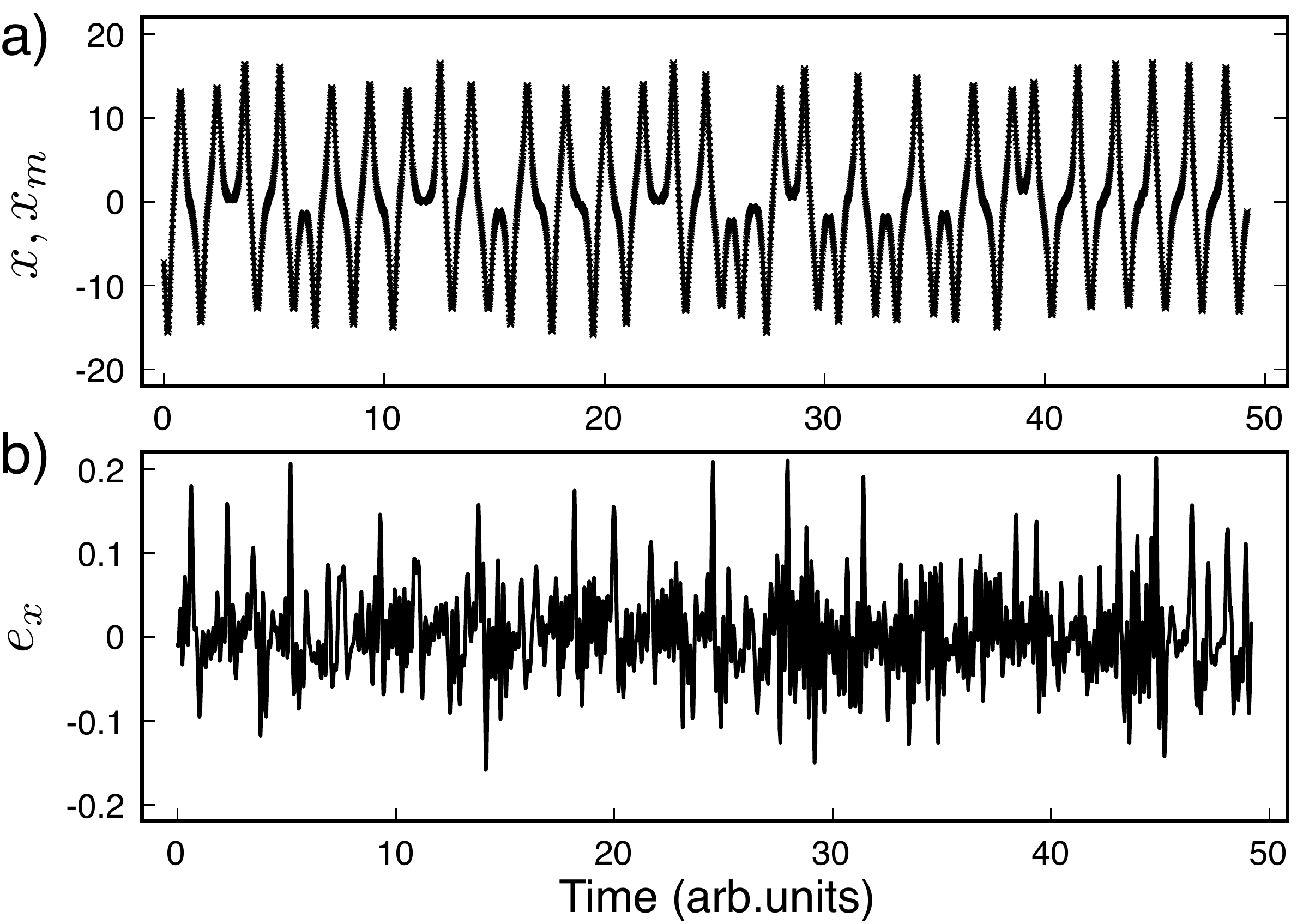}
\caption{{(a)} Experimental time-series (solid line) and model output (crosses). {(b)}  Difference between model and experiment. }
\label{fig:sync}
\end{figure}

We now discuss results of coupling the chaotic data, part of which is shown in Fig.~\ref{fig:ts}b, to the Lorenz model given by \eq{lorm}. When utilizing the provably sufficient coupling strength of $K= 2260$ suggested by \eq{couplstr}, we do find high-quality synchronization. What is more, we find that the model synchronizes to the data for much lower coupling strengths. An example is shown in Fig.~\ref{fig:sync}a, where we display experimental data and model output for a coupling strength of $K=100$ and find that they are indistinguishable to the eye. The difference $e_x=x_m - x$, shown in Fig.~\ref{fig:sync}b, has a magnitude that is about 1\% of the oscillation amplitude (note the difference in scale) and is of much higher frequency than the natural timescale of the chaos, consistent with the interpretation that noise in the data is the main source of the remaining error. 

The fact that synchronization takes place for the sufficient coupling strength of the proof and, as it turns out, for much smaller coupling strengths means that our wheel passes the test, it is not inconsistent with the Lorenz model. Indeed, within the limitations of the experiment, the presented data provides strong evidence that the Lorenz model is a good description of the water wheel dynamics.

\section{Discussion\label{sec:discussion}}

We have provided details on our inexpensive and easy to implement realization of the Malkus water wheel and demonstrated that it produces chaotic and periodic behavior as predicted theoretically.  We also showed that a numeric model implementing the Lorenz equations synchronizes to the wheel's chaotic motion with high fidelity. 
We discussed how chaos synchronization provides a test that, when not satisfied, can falsify the claim that the wheel is described by the Lorenz model.  Within the limitations of an experimental system, our wheel passes this test. 

No model of a real system can be perfect and noise is present in any measurement. It is therefore not surprising that we find  small differences between the data and model. The interesting question and open challenge is to develop tools that allow one to determine the main source of the remaining error.  

Perfect chaos synchronization between two (noise-free) chaotic systems is only achieved if the master system (the wheel) and the driven system (the simulation implementing \eq{lorm}) are exactly identical. One might therefore suspect that the parameters that were used in simulations were not exactly those corresponding to the wheel's operating condition because 
 there undoubtably is uncertainty in the experimentally determined parameters that are used to map the measured angle $\theta(t)$ and wheel operating conditions onto the variables and parameters of the Lorenz system.
Stated differently, one is confronted with the problem of finding optimal parameter values for a chaotic system based solely on ones knowledge of a single scalar time series, in this case $\theta(t)$. This is a difficult problem, in general, due to the sensitive dependence of chaotic systems on the parameters and initial conditions of the unknown state variables ($y$ and $z$). Recently proposed techniques, that are inspired by control engineering approaches, are in some cases able to solve this problem.\cite{Andrievskii2007,Yu2008,Sorrentino2009,Abarbanel2010} Utilizing these ideas, we designed an adaptive observer\cite{Andrievskii2007} that generates for a given time series an (optimal) estimate of the corresponding model parameter values. It is these optimized values that were used for all simulation results shown in this paper. A detailed discussion of the adaptive observer technique is beyond the scope of this paper. However, we note that, quite reassuringly, the optimized parameter values are in close correspondence to the estimates based on experimental measurements. For the chaotic time series the optimized values are $\sigma \approx 2.5$ and $\rho \approx 66$ as compared to the experimentally determined values of $\sigma \approx 2.7$ and $\rho \approx 69$. As a matter of fact, given the measurement uncertainties we find this close correspondence rather surprising.

In summary, of the three potential causes for the small differences between the data and model, namely noise, parameter mismatches, and structural insufficiencies of the model, we can exclude parameter mismatches. 
We know for certain that the data is noisy and therefore contributes to the observed differences. We cannot with certainty exclude model insufficiencies, but if they exist their effect is small. Thus, we think that the wheel described in this paper comes close to an ideal one, one whose dynamics is exactly described by the Lorenz equations. It can therefore be used to explore other interesting aspects of the Lorenz system, such as its bifurcations.

\begin{acknowledgments}
  This work was supported by the Research Corporation for Science Advancement (Award No. 7847).
We thank  Greg Eibel for helping with the construction of the wheel, Jay Ewing for help with the video analysis, and
 Adarsh Pyrelal for initial tests aimed at characterizing the wheel parameters. L.I. thanks the nonlinear dynamics group at Redstone Arsenal, Huntsville for their hospitality and inspiring conversations. 
 \end{acknowledgments}

\appendix

 \section{Numeric calculation of the Lyapunov Exponents}

For completeness we will recall in this section the definition of  Lyapunov exponents and provide some details on how to compute them numerically. Some reviews on this topic are found in Wolf et al.,\cite{Wolf1985}, Geist et al.,\cite{Geist1990} and Souza-Machado et al.\cite{Souza-Machado1990}

The object of study are nonlinear ordinary differential equations,
\begin{align}\label{ODE}
\dot{\x} &= \f(\x), \qquad \x(0)=\x_0,
\end{align}
where $\x(t)=(x_1(t),x_2(t),\ldots,x_n(t)) \in \Reals^n$, $\x_0$ is the initial condition, and $\f$ is an $n$-dimensional vector field. 
To define Lyapunov exponents associated with solutions $\x(t)$, one needs to take into account that the rate of separation of two trajectories in phase space can be different for different orientations of the initial separation vector. Therefore, one has to consider a spectrum of $n$ Lyapunov exponents for an $n$ dimensional phase space, $\{\lambda_1,\lambda_2,\ldots,\lambda_n\}$.  
Since the Lyapunov spectrum characterizes the evolution of infinitesimal perturbations, one can utilize a linearization around a fiduciary trajectory via solutions to the matrix differential equation 
\begin{align}\label{Yode}
\dot{\Y} &= \J \; \Y;     \qquad  \Y(\x_0,t=0)=\openone.
\end{align}
In this equation, $\J$ is the time-dependent Jacobian matrix with elements $ \J_{ij}[\x(t)] = (\partial f_i / \partial x_j)|_{\x=\x(t)}$ that are the partial derivatives of the vector field $\f$ in \eq{ODE} evaluated along the fiduciary trajectory $\x(t)$.
With the identity matrix as initial condition the  $n\times n$ matrix $\Y$ gives then the complete linearized flow map with respect to the standard orthonormal basis $\{\eh_1,\ldots,\eh_n\}$. In other words, $\Y(\x_0,t)$  describes the evolution in time of  both the magnitude and the orientation in phase space of any initial infinitesimal perturbation from $\x_0$ because $\delta \x(t) = \Y(\x_0,t) \delta \x_0$. The Lyapunov exponents $\lambda_i$ are defined by the logarithms of the (real) eigenvalues $\mu_i$ of the positive and symmetric matrix
\begin{align}\label{Lambda}
\Lambda = \lim_{t \to \infty} \left[ \Y^{\text{T}}(\x_0,t) \, \Y(\x_0,t) \right]^{1/{2t}},
\end{align}
where  $\Y^{\text{T}}(\x_0,t)$ denotes the transpose of $\Y(\x_0,t)$.
This implies that for every initial condition $\x_0$ there exists an orthonormal set of initial perturbation vectors $\delta \v_i$ such that 
\begin{align}\label{le}
\lambda_i &= \lim_{t \to \infty} \frac{1}{t}\, \ln\left|  \Y(\x_0,t) \, \delta \v_i\right|, & i&=1,2,\ldots,n.
\end{align}
 It was shown by Oseledets\cite{Oseledets1968} that the limits on the right-hand side of \eq{Lambda} and \eq{le} exist for almost every initial condition $\x_0$ and it has been argued\cite{Farmer1983} that for ergodic systems the values of the Lyapunov exponents $\{\lambda_i\}$ do not depend on the initial conditions (up to a measure zero in phase space). Thus, the Lyapunov exponents are global properties of the attractor of the dynamical system.

 The definition of the Lyapunov exponents might suggest to integrate the linearized equation~(\ref{Yode}) along with the nonlinear differential equation~(\ref{ODE}). This is not feasible because the exponential growth and decay rates of initial perturbation vectors quickly exceed the abilities of numerical number representation in a computer. Furthermore, perturbation vectors quickly align with the direction of maximal growth making it impossible to determine any but the largest Lyapunov exponent.   To compute the Lyapunov spectrum, the most commonly used algorithms are based on a step-wise procedure where \eq{Yode} is integrated for short intervals of time and the vectors spanning the phase space volume are reorthonormalized at each step. 
 
 To be precise, we compute the largest $k$ Lyapunov exponents by considering the evolution of the volume $V_k$ of $k$-dimensional parallelepipeds in $n$-dimensional phase space. It can be shown that $V_k$ is given in terms of the sum of the largest $k$ Lyapunov exponents\cite{Benettin1980,Benettin1980a}  via
 \begin{align}\label{sumleqr}
\sum_{i=1}^k \lambda_k = \lim_{t \to \infty}  \frac{1}{t}\, \ln\left[ V_k \right] \qquad (1\le k \le n)
\end{align}
 when the $\lambda_i$ form a monotonically decreasing sequence. Consider therefore an initial $k$-dimensional hypercube centered on the initial point $\x_0$, where the cube is defined via an orthogonal $n \times k$ matrix $\Pmat_0 = \left( \oh_1,\ldots,\oh_k\right)$, the columns of which are $k$ (randomly chosen) orthonormal $n$-dimensional vectors $\oh_i$ forming the cube's axes. Under the action of the flow the cube will deform into an parallelepiped $\Pmat(t) = \Y(\x_0,t) \Pmat(0)$ with volume $V_k$.  The volumes $V_k$ can be computed via the QR decomposition that factorizes a $n \times k$ matrix $\Pmat$
 \begin{align}
\Pmat = \Qmat \Rmat
\end{align} 
into the product of an orthogonal $n \times k$ matrix $\Qmat$ and an upper triangular $k \times k$ matrix $\Rmat$ with positive diagonal elements $R_{ii}$. The volume of $\Pmat$ is
\begin{align}
V_k = |\det \Pmat|  = |\det \Rmat| = \prod_{i=1}^k R_{ii}
\end{align}
Substituting this expression into \eq{sumleqr} for $V_k$ for all $k$ with $1 \le k \le n$, we find that
\begin{align}\label{LE_Rii}
\lambda_i =  \lim_{t \to \infty}  \frac{1}{t}\, \ln\left[ R_{ii}(t) \right] \qquad (i=1,\ldots, k)
\end{align}

We can calculate the time average of the $R_{ii}$ in \eq{LE_Rii} in a stable manner. To do so, first note that the evolution in time of $\Pmat$ is given by 
\begin{align}
\frac{d \Pmat}{dt} = \frac{d\Y}{dt} \, \Pmat(0) = \J \Pmat(t).
\end{align}
The advantage of considering the time evolution of $\Pmat$ is that the determination of the largest $k$ Lypunov exponents
involves the integration of just $n\,(1+k)$ differential equations instead of the $n\,(1+n)$ equations that would be necessary if one considered the evolution of $\Y$. The computation proceeds stepwise, where one integrates over a sufficiently small time-interval, $\Delta t_j = t_j - t_{j-1}$, the coupled differential equations 
\begin{align}
\begin{aligned}
\frac{d \x}{dt} &=\f(\x),  &   \x(t_{j-1}^+) &= \x(t_{j-1}^-) \\
\frac{d \Pmat}{dt} & = \J[\x(t)] \;  \Pmat,  &  \Pmat(t_{j-1}^+) &= \Qmat_{j-1},
\end{aligned}
&& t_{j-1} \le t \le t_j.
\end{align}
The initial conditions for the first step are $\x(0)=\x_0$ and $\Pmat(0)=\Pmat_0$. The size of the time step $\Delta t_j$ has to be chosen such that $\Pmat(t)$ remains well conditioned. Time steps on the order of a typical oscillation period are often suggested.  Next, the QR decomposition of the matrix $\Pmat_{j}=\Pmat(t_j)$, 
\begin{align}
\Pmat_j = \Qmat_j \, \Rmat_{j},
\end{align}
is performed to reorthonormalize the phase space volume.
The orthogonal matrix $\Qmat_j$ is used to initialize the integration of the next step and the diagonal matrix elements of  $\Rmat_{j}$ are accumulated to compute the Lyapunov exponents.
Since the  flow matrix $\Pmat=\Pmat_{m}\cdots\Pmat_0$ can be expressed as the product of the matrices $\Pmat_{j}$ computed at successive points along the orbit, $\x(t_j)$, above procedure implies that
\begin{align}
\Pmat=\Qmat_m \, \Rmat_{m} \, \Rmat_{m-1} \cdots  \, \Rmat_{1}.
\end{align}
Thus, the Lyapunov exponents are 
\begin{align}\label{lefinal}
\lambda_i = \lim_{m \to \infty} \frac{\sum_{j=1}^{m} \ln[R_{ii,j}]}{\sum_{j=1}^{m} \Delta t_j}.
\end{align}

\begin{figure}[htbp]
\centering
\includegraphics[width= \breite \columnwidth]{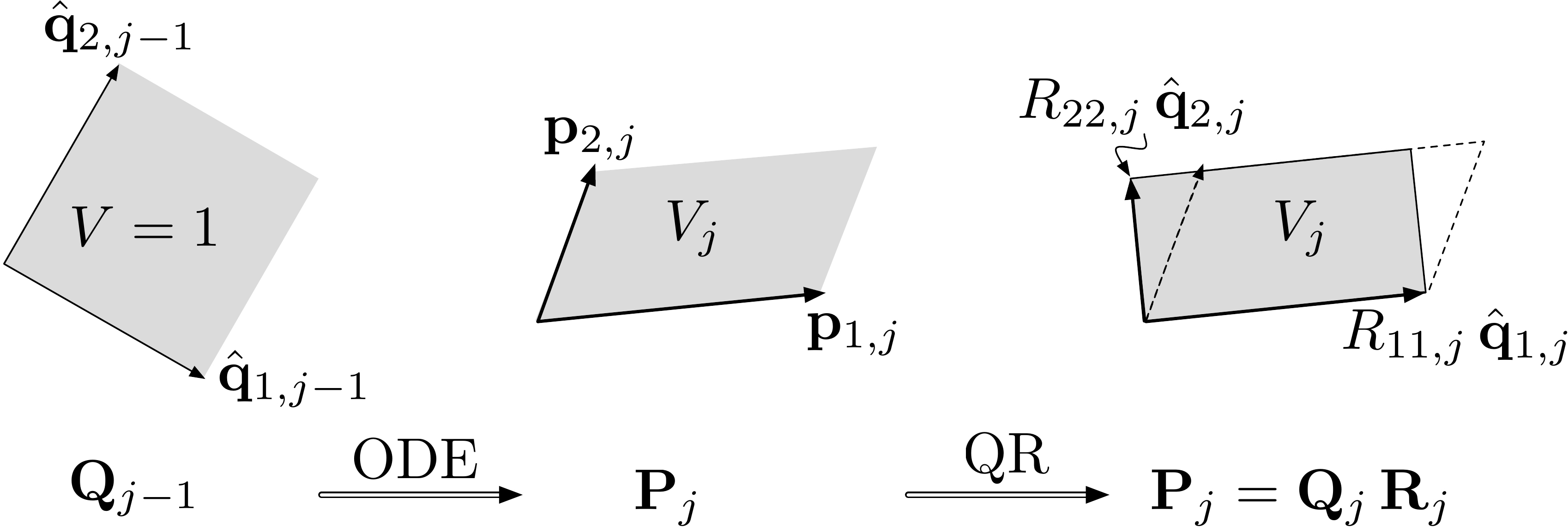}
\caption{ Geometric illustration of the QR decomposition scheme for computing Lyapunov exponents. }
\label{ fig:qr}
\end{figure}

We implement the above procedure using an adaptive stepsize integrator (the DDRIV fortran-integrator) and the QR decomposition routine contained in the LAPACK library. 

In general the convergence to the limit in \eq{lefinal} is very slow and it is useful to have checks on how well the exponents have been approximated. To do so, we note that for a dissipative systems, such as the chaotic Lorenz system, the total phase space volume contracts (exponentially in time on average), implying that the sum of all Lyapunov exponents, 
\begin{align}\label{sumle}
\sum_{i=1}^{3} \lambda_i = \lim_{t \to \infty}  \frac{1}{t}\, \ln\left| \text{det} \Y\right|= \lim_{t \to \infty}  \frac{1}{t}\, \ln\left[  V_3(t) \right],
\end{align}
must be negative.
For the Lorenz system with $b=1$ [\eq{lor}] with corresponding linearized flow equation
\begin{align}\label{lorjac}
\dot{\Y} &= \J \, \Y & \J &=\begin{pmatrix} -\sigma & \sigma & 0 \\ \rho-x_3 & -1 & -x_1\\ x_2 & x_1 & -1 \end{pmatrix}
\end{align}
  one can calculate the phase space volume contraction rate explicitly by applying  to equation \eq{lorjac} Liouville's formula
 \begin{align}
    \det\Y(\x_0,t)=\det\Y(\x_0,0)\,\exp\biggl(\int_{0}^t \mathrm{tr}\left[\J(\xi)\right] \,\textrm{d}\xi\biggr).
\end{align}
Since $\det \Y(\x_0,0) = \det \openone = 1$ and the trace of the Jacobian  is a constant, $\text{tr}[\J]=-\sigma-2$, \eq{sumle} implies that
\begin{align}\label{lorsumle}
\sum_{i=1}^{3} \lambda_i 
=  \lim_{t \to \infty}  \frac{1}{t}\,\int_{0}^t \mathrm{tr}\left[\J(\xi)\right] \,\textrm{d}\xi 
= -\sigma-2.
\end{align}
This provides a cross-check for convergence. A second check is obtained by noting that perturbations along a trajectory will neither grow nor shrink exponentially in time, implying that there is a zero Lyapunov exponent. This zero exponent exists whenever the attractor is not a fixed point, e.g. for periodic or chaotic dynamics. 

We utilize these checks in our numerical procedure by computing all three Lyapunov exponents for the Lorenz system and accepting the computed values whenever (a) the zero exponent, if it exists, has converged such that its magnitude is smaller than $10^{-3}$ and (b) the sum condition \eq{lorsumle} is satisfied, $\left|  -(2+\sigma)- \sum_{i=1}^3 \lambda_i \right| < 10^{-3}$.


\begin{thebibliography}{36}
\expandafter\ifx\csname natexlab\endcsname\relax\def\natexlab#1{#1}\fi
\expandafter\ifx\csname bibnamefont\endcsname\relax
  \def\bibnamefont#1{#1}\fi
\expandafter\ifx\csname bibfnamefont\endcsname\relax
  \def\bibfnamefont#1{#1}\fi
\expandafter\ifx\csname citenamefont\endcsname\relax
  \def\citenamefont#1{#1}\fi
\expandafter\ifx\csname url\endcsname\relax
  \def\url#1{\texttt{#1}}\fi
\expandafter\ifx\csname urlprefix\endcsname\relax\def\urlprefix{URL }\fi
\providecommand{\bibinfo}[2]{#2}
\providecommand{\eprint}[2][]{\url{#2}}

\bibitem[{\citenamefont{Lorenz}(1963)}]{Lorenz1963}
\bibinfo{author}{\bibfnamefont{E.~N.} \bibnamefont{Lorenz}},
  {``}\bibinfo{title}{Deterministic Nonperiodic Flow},{''} \bibinfo{journal}{J.
  Atmos. Sci.} \textbf{\bibinfo{volume}{20}}, 130-141 (\bibinfo{year}{1963}).

\bibitem[{\citenamefont{Dreyer and Hickey}(1991)}]{Dreyer1991}
\bibinfo{author}{\bibfnamefont{K.}~\bibnamefont{Dreyer}} \bibnamefont{and}
  \bibinfo{author}{\bibfnamefont{F.~R.} \bibnamefont{Hickey}},
  {``}\bibinfo{title}{The route to chaos in a dripping water faucet},{''}
  \bibinfo{journal}{Am. J. Phys.} \textbf{\bibinfo{volume}{59}}, 619--627
  (\bibinfo{year}{1991}).

\bibitem[{\citenamefont{Corron et~al.}(2000)\citenamefont{Corron, Pethel, and
  Hopper}}]{Corron2000}
\bibinfo{author}{\bibfnamefont{N.~J.} \bibnamefont{Corron}},
  \bibinfo{author}{\bibfnamefont{S.~D.} \bibnamefont{Pethel}},
  \bibnamefont{and} \bibinfo{author}{\bibfnamefont{B.~A.}
  \bibnamefont{Hopper}}, {``}\bibinfo{title}{Controlling Chaos with Simple
  Limiters},{''} \bibinfo{journal}{Phys. Rev. Lett.}
  \textbf{\bibinfo{volume}{84}}, 3835--3838 (\bibinfo{year}{2000}).

\bibitem[{\citenamefont{Schmitz et~al.}(1977)\citenamefont{Schmitz, Graziani,
  and Hudson}}]{Schmitz1977}
\bibinfo{author}{\bibfnamefont{R.~A.} \bibnamefont{Schmitz}},
  \bibinfo{author}{\bibfnamefont{K.~R.} \bibnamefont{Graziani}},
  \bibnamefont{and} \bibinfo{author}{\bibfnamefont{J.~L.}
  \bibnamefont{Hudson}}, {``}\bibinfo{title}{Experimental Evidence of Chaotic
  States in the Belousov-Zhabotinskii Reaction},{''} \bibinfo{journal}{J. Chem.
  Phys.} \textbf{\bibinfo{volume}{67}}, 3040-3044 (\bibinfo{year}{1977}).

\bibitem[{\citenamefont{Illing et~al.}(2007)\citenamefont{Illing, Gauthier, and
  Roy}}]{Illing2007}
\bibinfo{author}{\bibfnamefont{L.}~\bibnamefont{Illing}},
  \bibinfo{author}{\bibfnamefont{D.~J.} \bibnamefont{Gauthier}},
  \bibnamefont{and} \bibinfo{author}{\bibfnamefont{R.}~\bibnamefont{Roy}}, in
  \emph{\bibinfo{booktitle}{Advances in Atomic, Molecular, and Optical
  Physics}}, edited by \bibinfo{editor}{\bibfnamefont{P.~R.}
  \bibnamefont{Berman}},
  \bibinfo{editor}{\bibfnamefont{E.}~\bibnamefont{Arimondo}}, \bibnamefont{and}
  \bibinfo{editor}{\bibfnamefont{C.}~\bibnamefont{Lin}}
  (\bibinfo{publisher}{Elsevier, Amsterdam}, \bibinfo{year}{2007}),
  vol.~\bibinfo{volume}{54}, pp. \bibinfo{pages}{615--695}.

\bibitem[{\citenamefont{Argyris et~al.}(2005)\citenamefont{Argyris, Syvridis,
  Larger, Annovazzi-Lodi, Colet, Fischer, Garcia-Ojalvo, Mirasso, Pesquera, and
  Shore}}]{Argyris2005}
\bibinfo{author}{\bibfnamefont{A.}~\bibnamefont{Argyris}},
  \bibinfo{author}{\bibfnamefont{D.}~\bibnamefont{Syvridis}},
  \bibinfo{author}{\bibfnamefont{L.}~\bibnamefont{Larger}},
  \bibinfo{author}{\bibfnamefont{V.}~\bibnamefont{Annovazzi-Lodi}},
  \bibinfo{author}{\bibfnamefont{P.}~\bibnamefont{Colet}},
  \bibinfo{author}{\bibfnamefont{I.}~\bibnamefont{Fischer}},
  \bibinfo{author}{\bibfnamefont{J.}~\bibnamefont{Garcia-Ojalvo}},
  \bibinfo{author}{\bibfnamefont{C.~R.} \bibnamefont{Mirasso}},
  \bibinfo{author}{\bibfnamefont{L.}~\bibnamefont{Pesquera}}, \bibnamefont{and}
  \bibinfo{author}{\bibfnamefont{K.~A.} \bibnamefont{Shore}},
  {``}\bibinfo{title}{Chaos-based communications at high bit rates using
  commercial fibre-optic links},{''} \bibinfo{journal}{Nature}
  \textbf{\bibinfo{volume}{438}}, 343--346 (\bibinfo{year}{2005}).

\bibitem[{\citenamefont{Cavalcante et~al.}(2010)\citenamefont{Cavalcante,
  Gauthier, Socolar, and Zhang}}]{Cavalcante2010}
\bibinfo{author}{\bibfnamefont{H.~L. D. d.~S.} \bibnamefont{Cavalcante}},
  \bibinfo{author}{\bibfnamefont{D.~J.} \bibnamefont{Gauthier}},
  \bibinfo{author}{\bibfnamefont{J.~E.~S.} \bibnamefont{Socolar}},
  \bibnamefont{and} \bibinfo{author}{\bibfnamefont{R.}~\bibnamefont{Zhang}},
  {``}\bibinfo{title}{On the origin of chaos in autonomous
  BooleanÊnetworks},{''} \bibinfo{journal}{Phil. Trans. R. Soc. A}
  \textbf{\bibinfo{volume}{368}}, 495--513 (\bibinfo{year}{2010}).

\bibitem[{\citenamefont{Callan et~al.}(2010)\citenamefont{Callan, Illing, Gao,
  Gauthier, and Sch\"{o}ll}}]{Callan2010}
\bibinfo{author}{\bibfnamefont{K.~E.} \bibnamefont{Callan}},
  \bibinfo{author}{\bibfnamefont{L.}~\bibnamefont{Illing}},
  \bibinfo{author}{\bibfnamefont{Z.}~\bibnamefont{Gao}},
  \bibinfo{author}{\bibfnamefont{D.~J.} \bibnamefont{Gauthier}},
  \bibnamefont{and}
  \bibinfo{author}{\bibfnamefont{E.}~\bibnamefont{Sch\"{o}ll}},
  {``}\bibinfo{title}{Broadband Chaos Generated by an Optoelectronic
  Oscillator},{''} \bibinfo{journal}{Phys. Rev. Lett.}
  \textbf{\bibinfo{volume}{104}}, 113901-4 (\bibinfo{year}{2010}).

\bibitem[{\citenamefont{Lorenz}(1993)}]{Lorenz1993}
\bibinfo{author}{\bibfnamefont{E.~N.} \bibnamefont{Lorenz}},
  \emph{\bibinfo{title}{The Essence of Chaos}} (\bibinfo{publisher}{University
  of Washington Press, Seattle, WA}, \bibinfo{year}{1993}).

\bibitem[{\citenamefont{Malkus}(1972)}]{Malkus1972}
\bibinfo{author}{\bibfnamefont{W.~V.~R.} \bibnamefont{Malkus}},
  {``}\bibinfo{title}{Non-periodic convection at high and low Prandtl
  number},{''} \bibinfo{journal}{Mem. Soc. R. Sci. Liege Collect.}
  \textbf{\bibinfo{volume}{IV}}, 125--128 (\bibinfo{year}{1972}).

\bibitem[{\citenamefont{Strogatz}(1994)}]{Strogatz1994}
\bibinfo{author}{\bibfnamefont{S.~H.} \bibnamefont{Strogatz}},
  \emph{\bibinfo{title}{Nonlinear Dynamics And Chaos}}
  (\bibinfo{publisher}{Reading, MA: Addison-Wesley}, \bibinfo{year}{1994}).

\bibitem[{\citenamefont{Tylee}(1995)}]{Tylee1995}
\bibinfo{author}{\bibfnamefont{J.~L.} \bibnamefont{Tylee}},
  {``}\bibinfo{title}{Chaos in a Real System},{''}
  \bibinfo{journal}{Simulation} \textbf{\bibinfo{volume}{64}}, 176--183
  (\bibinfo{year}{1995}).

\bibitem[{\citenamefont{Wiederick et~al.}(1987)\citenamefont{Wiederick,
  Gauthier, Campbell, and Rochon}}]{Wiederick1987}
\bibinfo{author}{\bibfnamefont{H.~D.} \bibnamefont{Wiederick}},
  \bibinfo{author}{\bibfnamefont{N.}~\bibnamefont{Gauthier}},
  \bibinfo{author}{\bibfnamefont{D.~A.} \bibnamefont{Campbell}},
  \bibnamefont{and} \bibinfo{author}{\bibfnamefont{P.}~\bibnamefont{Rochon}},
  {``}\bibinfo{title}{Magnetic braking: Simple theory and experiment},{''}
  \bibinfo{journal}{Am. J. Phys.} \textbf{\bibinfo{volume}{55}}, 500--503
  (\bibinfo{year}{1987}).

\bibitem[{\citenamefont{Heald}(1988)}]{Heald1988}
\bibinfo{author}{\bibfnamefont{M.~A.} \bibnamefont{Heald}},
  {``}\bibinfo{title}{Magnetic braking: Improved theory},{''}
  \bibinfo{journal}{Am. J. Phys.} \textbf{\bibinfo{volume}{56}}, 521--522
  (\bibinfo{year}{1988}).

\bibitem[{\citenamefont{Marcuso et~al.}(1991)\citenamefont{Marcuso, Gass,
  Jones, and Rowlett}}]{Marcuso1991a}
\bibinfo{author}{\bibfnamefont{M.}~\bibnamefont{Marcuso}},
  \bibinfo{author}{\bibfnamefont{R.}~\bibnamefont{Gass}},
  \bibinfo{author}{\bibfnamefont{D.}~\bibnamefont{Jones}}, \bibnamefont{and}
  \bibinfo{author}{\bibfnamefont{C.}~\bibnamefont{Rowlett}},
  {``}\bibinfo{title}{Magnetic drag in the quasi-static limit: A computational
  method},{''} \bibinfo{journal}{Am. J. Phys.} \textbf{\bibinfo{volume}{59}},
  1118--1123 (\bibinfo{year}{1991}).

\bibitem[{\citenamefont{Matson}(2007)}]{Matson2007}
\bibinfo{author}{\bibfnamefont{L.~E.} \bibnamefont{Matson}},
  {``}\bibinfo{title}{The Malkus--Lorenz water wheel revisited},{''}
  \bibinfo{journal}{Am. J. Phys.} \textbf{\bibinfo{volume}{75}}, 1114--1122
  (\bibinfo{year}{2007}).

\bibitem[{\citenamefont{Pnueli and Gutfinger}(1992)}]{Pnueli1992}
\bibinfo{author}{\bibfnamefont{D.}~\bibnamefont{Pnueli}} \bibnamefont{and}
  \bibinfo{author}{\bibfnamefont{C.}~\bibnamefont{Gutfinger}},
  \emph{\bibinfo{title}{Fluid Mechanics}} (\bibinfo{publisher}{Cambridge
  University Press}, \bibinfo{year}{1992}).

\bibitem[{\citenamefont{Sparrow}(1982)}]{Sparrow1982}
\bibinfo{author}{\bibfnamefont{C.}~\bibnamefont{Sparrow}},
  \emph{\bibinfo{title}{The Lorenz Equations: Bifurcations, Chaos and Strange
  Attractors}} (\bibinfo{publisher}{Springer-Verlag, New York},
  \bibinfo{year}{1982}).

\bibitem[{\citenamefont{Galias and Zgliczynski}(1998)}]{Galias1998}
\bibinfo{author}{\bibfnamefont{Z.}~\bibnamefont{Galias}} \bibnamefont{and}
  \bibinfo{author}{\bibfnamefont{P.}~\bibnamefont{Zgliczynski}},
  {``}\bibinfo{title}{Computer assisted proof of chaos in the Lorenz
  equations},{''} \bibinfo{journal}{Physica D} \textbf{\bibinfo{volume}{115}},
  165--188 (\bibinfo{year}{1998}).

\bibitem[{\citenamefont{Tucker}(1999)}]{Tucker1999}
\bibinfo{author}{\bibfnamefont{W.}~\bibnamefont{Tucker}},
  {``}\bibinfo{title}{The Lorenz attractor exists},{''} \bibinfo{journal}{C. R.
  Acad. Sci. Paris Ser. I Math.} \textbf{\bibinfo{volume}{328}}, 1197-1202
  (\bibinfo{year}{1999}).

\bibitem[{\citenamefont{Mischaikow et~al.}(2001)\citenamefont{Mischaikow,
  Mrozek, and Szymczak}}]{Mischaikow2001}
\bibinfo{author}{\bibfnamefont{K.}~\bibnamefont{Mischaikow}},
  \bibinfo{author}{\bibfnamefont{M.}~\bibnamefont{Mrozek}}, \bibnamefont{and}
  \bibinfo{author}{\bibfnamefont{A.}~\bibnamefont{Szymczak}},
  {``}\bibinfo{title}{Chaos in the Lorenz Equations: A Computer Assisted Proof
  Part III: Classical Parameter Values},{''} \bibinfo{journal}{J. Diff. Eqs.}
  \textbf{\bibinfo{volume}{169}}, 17--56 (\bibinfo{year}{2001}).

\bibitem[{\citenamefont{Abarbanel}(1995)}]{Abarbanel1995}
\bibinfo{author}{\bibfnamefont{H.~D.~I.} \bibnamefont{Abarbanel}},
  \emph{\bibinfo{title}{Analysis of Observed Chaotic Data}}
  (\bibinfo{publisher}{Springer, New York}, \bibinfo{year}{1995}).

\bibitem[{\citenamefont{Kantz and Schreiber}(2004)}]{Kantz2004}
\bibinfo{author}{\bibfnamefont{H.}~\bibnamefont{Kantz}} \bibnamefont{and}
  \bibinfo{author}{\bibfnamefont{T.}~\bibnamefont{Schreiber}},
  \emph{\bibinfo{title}{Nonlinear Time Series Analysis}}
  (\bibinfo{publisher}{Cambridge University Press}, \bibinfo{year}{2004}).

\bibitem[{\citenamefont{Pikovsky et~al.}(2003)\citenamefont{Pikovsky,
  Rosenblum, and Kurths}}]{Pikovsky2003}
\bibinfo{author}{\bibfnamefont{A.}~\bibnamefont{Pikovsky}},
  \bibinfo{author}{\bibfnamefont{M.}~\bibnamefont{Rosenblum}},
  \bibnamefont{and} \bibinfo{author}{\bibfnamefont{J.}~\bibnamefont{Kurths}},
  \emph{\bibinfo{title}{Synchronization: a universal concept in nonlinear
  sciences}} (\bibinfo{publisher}{Cambridge University Press, Cambridge},
  \bibinfo{year}{2003}).

\bibitem[{\citenamefont{Swinnerton-Dyer}(2001)}]{Swinnerton2001}
\bibinfo{author}{\bibfnamefont{P.}~\bibnamefont{Swinnerton-Dyer}},
  {``}\bibinfo{title}{Bounds for trajectories of the Lorenz equations: an
  illustration of how to choose Liapunov functions},{''}
  \bibinfo{journal}{Phys. Lett. A} \textbf{\bibinfo{volume}{281}}, 161--167
  (\bibinfo{year}{2001}).

\bibitem[{\citenamefont{Andrievskii et~al.}(2007)\citenamefont{Andrievskii,
  Nikiforov, and Fradkov}}]{Andrievskii2007}
\bibinfo{author}{\bibfnamefont{B.~R.} \bibnamefont{Andrievskii}},
  \bibinfo{author}{\bibfnamefont{V.~O.} \bibnamefont{Nikiforov}},
  \bibnamefont{and} \bibinfo{author}{\bibfnamefont{A.~L.}
  \bibnamefont{Fradkov}}, {``}\bibinfo{title}{Adaptive observer-based
  synchronization of the nonlinear nonpassifiable systems},{''}
  \bibinfo{journal}{Automat. Rem. Contr.} \textbf{\bibinfo{volume}{68}},
  1186--1200 (\bibinfo{year}{2007}).

\bibitem[{\citenamefont{Yu and Parlitz}(2008)}]{Yu2008}
\bibinfo{author}{\bibfnamefont{D.~C.} \bibnamefont{Yu}} \bibnamefont{and}
  \bibinfo{author}{\bibfnamefont{U.}~\bibnamefont{Parlitz}},
  {``}\bibinfo{title}{Estimating parameters by autosynchronization with
  dynamics restrictions},{''} \bibinfo{journal}{Phys. Rev. E}
  \textbf{\bibinfo{volume}{77}}, 066221-7 (\bibinfo{year}{2008}).

\bibitem[{\citenamefont{Sorrentino and Ott}(2009)}]{Sorrentino2009}
\bibinfo{author}{\bibfnamefont{F.}~\bibnamefont{Sorrentino}} \bibnamefont{and}
  \bibinfo{author}{\bibfnamefont{E.}~\bibnamefont{Ott}},
  {``}\bibinfo{title}{Using synchronization of chaos to identify the dynamics
  of unknown systems},{''} \bibinfo{journal}{Chaos}
  \textbf{\bibinfo{volume}{19}}, 033108-8 (\bibinfo{year}{2009}).

\bibitem[{\citenamefont{Abarbanel et~al.}(2010)\citenamefont{Abarbanel, Kostuk,
  and Whartenby}}]{Abarbanel2010}
\bibinfo{author}{\bibfnamefont{H.~D.~I.} \bibnamefont{Abarbanel}},
  \bibinfo{author}{\bibfnamefont{M.}~\bibnamefont{Kostuk}}, \bibnamefont{and}
  \bibinfo{author}{\bibfnamefont{W.}~\bibnamefont{Whartenby}},
  {``}\bibinfo{title}{Data assimilation with regularized nonlinear
  instabilities},{''} \bibinfo{journal}{Q. J. R. Meteorol. Soc.}
  \textbf{\bibinfo{volume}{136}}, 769--783 (\bibinfo{year}{2010}).

\bibitem[{\citenamefont{Wolf et~al.}(1985)\citenamefont{Wolf, Swift, Swinney,
  and Vastano}}]{Wolf1985}
\bibinfo{author}{\bibfnamefont{A.}~\bibnamefont{Wolf}},
  \bibinfo{author}{\bibfnamefont{J.~B.} \bibnamefont{Swift}},
  \bibinfo{author}{\bibfnamefont{H.~L.} \bibnamefont{Swinney}},
  \bibnamefont{and} \bibinfo{author}{\bibfnamefont{J.~A.}
  \bibnamefont{Vastano}}, {``}\bibinfo{title}{Determining Lyapunov Exponents
  From A Time-Series},{''} \bibinfo{journal}{Physica D}
  \textbf{\bibinfo{volume}{16}}, 285--317 (\bibinfo{year}{1985}).

\bibitem[{\citenamefont{Geist et~al.}(1990)\citenamefont{Geist, Parlitz, and
  Lauterborn}}]{Geist1990}
\bibinfo{author}{\bibfnamefont{K.}~\bibnamefont{Geist}},
  \bibinfo{author}{\bibfnamefont{U.}~\bibnamefont{Parlitz}}, \bibnamefont{and}
  \bibinfo{author}{\bibfnamefont{W.}~\bibnamefont{Lauterborn}},
  {``}\bibinfo{title}{Comparison Of Different Methods For Computing Lyapunov
  Exponents},{''} \bibinfo{journal}{Prog. Theor. Phys.}
  \textbf{\bibinfo{volume}{83}}, 875--893 (\bibinfo{year}{1990}).

\bibitem[{\citenamefont{Souza-Machado et~al.}(1990)\citenamefont{Souza-Machado,
  Rollins, Jacobs, and Hartman}}]{Souza-Machado1990}
\bibinfo{author}{\bibfnamefont{S.~D.} \bibnamefont{Souza-Machado}},
  \bibinfo{author}{\bibfnamefont{R.~W.} \bibnamefont{Rollins}},
  \bibinfo{author}{\bibfnamefont{D.~T.} \bibnamefont{Jacobs}},
  \bibnamefont{and} \bibinfo{author}{\bibfnamefont{J.~L.}
  \bibnamefont{Hartman}}, {``}\bibinfo{title}{Studying chaotic systems using
  microcomputer simulations and Lyapunov exponents},{''} \bibinfo{journal}{Am.
  J. Phys.} \textbf{\bibinfo{volume}{58}}, 321--329 (\bibinfo{year}{1990}).

\bibitem[{\citenamefont{Oseledets}(1968)}]{Oseledets1968}
\bibinfo{author}{\bibfnamefont{V.~I.} \bibnamefont{Oseledets}},
  {``}\bibinfo{title}{A multiplicative ergodic theorem. Lyapunov characteristic
  numbers for dynamical systems},{''} \bibinfo{journal}{Trans. Moscow Math.
  Soc.} \textbf{\bibinfo{volume}{19}}, 197-231 (\bibinfo{year}{1968}).

\bibitem[{\citenamefont{Farmer et~al.}(1983)\citenamefont{Farmer, Ott, and
  Yorke}}]{Farmer1983}
\bibinfo{author}{\bibfnamefont{J.~D.} \bibnamefont{Farmer}},
  \bibinfo{author}{\bibfnamefont{E.}~\bibnamefont{Ott}}, \bibnamefont{and}
  \bibinfo{author}{\bibfnamefont{J.~A.} \bibnamefont{Yorke}},
  {``}\bibinfo{title}{The Dimension Of Chaotic Attractors},{''}
  \bibinfo{journal}{Physica D} \textbf{\bibinfo{volume}{7}}, 153--180
  (\bibinfo{year}{1983}).

\bibitem[{\citenamefont{Benettin
  et~al.}(1980{\natexlab{a}})\citenamefont{Benettin, Galgani, Giorgilli, and
  Strelcyn}}]{Benettin1980}
\bibinfo{author}{\bibfnamefont{G.}~\bibnamefont{Benettin}},
  \bibinfo{author}{\bibfnamefont{L.}~\bibnamefont{Galgani}},
  \bibinfo{author}{\bibfnamefont{A.}~\bibnamefont{Giorgilli}},
  \bibnamefont{and} \bibinfo{author}{\bibfnamefont{J.~M.}
  \bibnamefont{Strelcyn}}, {``}\bibinfo{title}{Lyapunov characteristic
  exponents for smooth dynamical systems and for Hamiltonian systems; A method
  for computing all of them, Part 1: Theory},{''} \bibinfo{journal}{Meccanica}
  \textbf{\bibinfo{volume}{9}}, 9 - 20 (\bibinfo{year}{1980}{\natexlab{a}}).

\bibitem[{\citenamefont{Benettin
  et~al.}(1980{\natexlab{b}})\citenamefont{Benettin, Galgani, Giorgilli, and
  Strelcyn}}]{Benettin1980a}
\bibinfo{author}{\bibfnamefont{G.}~\bibnamefont{Benettin}},
  \bibinfo{author}{\bibfnamefont{L.}~\bibnamefont{Galgani}},
  \bibinfo{author}{\bibfnamefont{A.}~\bibnamefont{Giorgilli}},
  \bibnamefont{and} \bibinfo{author}{\bibfnamefont{J.~M.}
  \bibnamefont{Strelcyn}}, {``}\bibinfo{title}{Lyapunov characteristic
  exponents for smooth dynamical systems and for Hamiltonian systems; A method
  for computing all of them, Part 2: Numerical Application},{''}
  \bibinfo{journal}{Meccanica} \textbf{\bibinfo{volume}{9}}, 21-30
  (\bibinfo{year}{1980}{\natexlab{b}}).


\end{thebibliography}


\end{document}